\documentclass[fleqn,usenatbib]{mnras}

\usepackage{newtxtext,newtxmath}

\usepackage[T1]{fontenc}

\usepackage{graphicx}	
\usepackage{amsmath}	
\usepackage{amssymb}	
\usepackage{CJK} 
\usepackage{multirow} 

\newcommand{\ub}{\textit{u}'}
\newcommand{\gb}{\textit{g}'}
\newcommand{\rb}{\textit{r}'}
\newcommand{\ib}{\textit{i}'}
\newcommand{\Rb}{\textit{R}}
\newcommand{\Jb}{\textit{J}}
\newcommand{\Kb}{\textit{K}}
\newcommand{\Exopb}{\textit{ExoP}}

\title[WD\,1145+017: Bluing]{Once in a blue moon: detection of `bluing' during debris transits in the white dwarf WD\,1145+017}

\date{Accepted XXX. Received YYY; in original form ZZZ}

\pubyear{2017}

\begin{document}
\label{firstpage}
\pagerange{\pageref{firstpage}--\pageref{lastpage}}

\begin{CJK}{UTF8}{gbsn}
\author[N. Hallakoun et al.]{
N. Hallakoun,$^{1,2}$\thanks{E-mail: \href{mailto:naama@wise.tau.ac.il}{naama@wise.tau.ac.il}}
S.~Xu (许\CJKfamily{bsmi}偲\CJKfamily{gbsn}艺),$^{2}$
D. Maoz,$^{1}$
T.~R.~Marsh,$^{3}$
V.~D.~Ivanov,$^{4,2}$
\newauthor{
V.~S.~Dhillon,$^{5,6}$
M.~C.~P.~Bours,$^{7}$
S.~G.~Parsons,$^{5}$
P.~Kerry,$^{5}$
S.~Sharma,$^{8}$
K.~Su,$^{9}$
}
\newauthor{
S.~Rengaswamy,$^{10}$
P.~Pravec,$^{11}$
P.~Ku\v{s}nir\'ak,$^{11}$
H.~Ku\v{c}\'akov\'a,$^{12}$
J.~D.~Armstrong,$^{13,14}$
}
\newauthor{
C.~Arnold,$^{13}$
N.~Gerard$^{13}$
and L.~Vanzi$^{15}$}
\\
\\
$^{1}$School of Physics and Astronomy, Tel-Aviv University, Tel-Aviv 6997801, Israel\\
$^{2}$European Southern Observatory, Karl-Schwarzschild-Stra{\ss}e 2, D-85748 Garching, Germany\\
$^{3}$Department of Physics, University of Warwick, Coventry CV4 7AL, UK\\
$^{4}$European Southern Observatory, Ave. Alonso de C\'{o}rdova 3107, Vitacura, Santiago, Chile\\
$^{5}$Department of Physics and Astronomy, University of Sheffield, Sheffield S3 7RH, UK\\
$^{6}$Instituto de Astrof\'{i}sica de Canarias, E-38205 La Laguna, Santa Cruz de Tenerife, Spain\\
$^{7}$Departmento de F\'{i}sico y Astronom\'{i}a, Universidad de Valpara\'{i}so, Avenida Gran Breta\~{n}a 1111, Valpara\'{i}so, Chile\\
$^{8}$Aryabhatta Research Institute of Observational Sciences, Nainital 263001, India\\
$^{9}$Steward Observatory, University of Arizona, 933 North Cherry Avenue, Tucson, AZ 85721, USA\\
$^{10}$Indian Institute of Astrophysics, Koramangala 2nd Block, Bengaluru 560034, India\\
$^{11}$Astronomical Institute, Academy of Sciences of the Czech Republic, Fri\v{c}ova 1, 25165 Ond\v{r}ejov, Czech Republic\\
$^{12}$Astronomical Institute of the Charles University, Faculty of Mathemathics and Physics, V Hole\v{s}ovi\v{c}k\'{a}ch 2, 180 00 Praha 8, Czech Republic\\
$^{13}$University of Hawaii, Institute for Astronomy, 34 Ohia Ku Street, Pukalani, Hawaii 96768, USA\\
$^{14}$Las Cumbres Observatory Global Telescope Network, Inc. 6740 Cortona Drive Suite 102, Goleta, CA 93117, USA\\
$^{15}$\parbox[t]{\textwidth}{Department of Electrical Engineering and Center of Astro Engineering, Pontificia Universidad Cat\'{o}lica de Chile, Av. Vicu\~{n}a Mackenna 4860,\\ Santiago 7820436, Chile}
}

\maketitle

\begin{abstract}
The first transiting planetesimal orbiting a white dwarf was recently detected in \textit{K2} data of WD\,1145+017 and has been followed up intensively. The multiple, long, and variable transits suggest the transiting objects are dust clouds, probably produced by a disintegrating asteroid. In addition, the system contains circumstellar gas, evident by broad absorption lines, mostly in the \ub-band, and a dust disc, indicated by an infrared excess. Here we present the first detection of a change in colour of WD\,1145+017 during transits, using simultaneous multi-band fast-photometry ULTRACAM measurements over the \ub\gb\rb\ib-bands. The observations reveal what appears to be `bluing' during transits; transits are deeper in the redder bands, with a $\ub-\rb$ colour difference of up to $\sim -0.05$\,mag. We explore various possible explanations for the bluing. `Spectral' photometry obtained by integrating over bandpasses in the spectroscopic data in- and out-of-transit, compared to the photometric data, shows that the observed colour difference is most likely the result of reduced circumstellar absorption in the spectrum during transits. This indicates that the transiting objects and the gas share the same line-of-sight, and that the gas covers the white dwarf only partially, as would be expected if the gas, the transiting debris, and the dust emitting the infrared excess, are part of the same general disc structure (although possibly at different radii). In addition, we present the results of a week-long monitoring campaign of the system.
\end{abstract}

\begin{keywords}
stars: individual: WD\,1145+017 -- white dwarfs -- minor planets, asteroids: general -- techniques: photometric -- eclipses
\end{keywords}



\section{Introduction}
\label{sec:Intro}
Over 95~per cent of all stars will end their lives as white dwarfs \citep[WDs;][]{Althaus_2010}. As the majority of Sun-like stars host planets \citep[e.g.][]{Winn_2015, Shvartzvald_2016}, the fate of planetary systems can be studied by examining the immediate surroundings of WDs \citep{Veras_2016}. Recent studies show that about 25-50~per cent of all WDs exhibit `pollution' -- traces of heavy elements in their atmospheres \citep{Zuckerman_2003,Zuckerman_2010, Koester_2014}. The most heavily polluted WDs also have dust discs within their tidal radii, indicated by excess infrared (IR) radiation emitted by the dust, which feed the host WD with heavy elements \citep[e.g.][]{Kilic_2006}. Since the strong surface gravity of a WD causes all heavy elements to settle quickly below the photosphere, pollution in WDs cooler than $\sim 20,000$\,K is a strong indication for external accretion, likely from planetary debris \citep{Jura_2003, Koester_2014}. However, the object (or objects) supplying the accreting material, until recently, had not been directly observed.

Recently, the first direct evidence of a planetary-mass body orbiting a WD was found using data from the \textit{Kepler} extended mission (\textit{K2}) \citep{Vanderburg_2015}. The light curves of WD\,1145+017 acquired from \textit{K2} and from follow-up observations revealed multiple transit events with varying durations, depths, and shapes, interpreted to indicate the presence of a disintegrating asteroid orbiting the WD \citep{Vanderburg_2015, Croll_2017, Gaensicke_2016, Rappaport_2016, Alonso_2016, Zhou_2016, Gary_2017}. The transits exhibit several features which suggest that they are caused by dust clouds emitted by the asteroidal fragments, rather than by the fragments themselves: The transit durations ($3-50$\,min) are longer than expected for a solid object ($\sim 1$\,min); the shape of the transits is asymmetric; and the transit depths are variable and large ($\sim10-60$~per cent). Although the \textit{K2} light curve has shown six stable periods, ranging from $4.5$ to $5.0$ hours, the follow-up observations have detected only the shortest, $\sim 4.5$\,h, period (with the exception of \citealt{Gary_2017}, see Appendix~\ref{sec:LC}). Recent simulations suggest a differentiated parent asteroid orbiting within the tidal disruption radius for mantle-density material \citep{Veras_2017}. The light curve changes on a daily basis, completely altering its appearance after a few months, indicating the system is rapidly evolving.

High-resolution spectroscopic observations of the system revealed photospheric absorption lines from 11 heavy elements \citep{Xu_2016}, showing that the WD belongs to the `polluted' class that is actively accreting circumstellar material. The composition of the accreted material is similar to that of rocky objects within the solar system. Consistent with its heavily polluted atmosphere, this system also shows near-IR (NIR) excess from a dust disc \citep{Vanderburg_2015}. In addition, WD\,1145+017 is surrounded by circumstellar gas, evident by numerous, unusually broad ($\sim300$\,km\,s$^{-1}$), absorption lines in its spectrum \citep{Xu_2016}. The overall shape of the circumstellar lines changes on a monthly timescale and their depths can also change during transits \citep[][Xu et al., in preparation]{Redfield_2016}.

Since the transiting debris\footnote{By `debris' we mean the solid material, whether planetesimals or large dust grains, that produces the transits. The relation is still not clear between this material and the `dust disc' whose presence has been deduced from the NIR excess in this object \citep{Vanderburg_2015}.} may consist of particles small enough to be called `dust', and dust extinction depends on the grain-size to wavelength ratio and on the grain composition \citep{Bohren_1983}, simultaneous monitoring in different bands might reveal its properties. Previous studies have not found a significant dependence of transit depth on wavelength, thus constraining the grain size to $\gtrsim 0.8\,\mu$m \citep{Croll_2017, Alonso_2016, Zhou_2016}. In this paper we present the first detection of colour changes during transit in the light curve of WD\,1145+017, using multi-band fast-photometry obtained simultaneously in the \ub- \gb- and \rb/\ib-bands. Surprisingly, the light curves feature `bluing', rather than reddening, during transits (i.e. transits are deeper in the redder bands than in the \ub-band). The ULTRACAM observations are presented in Section~\ref{sec:Obs}. In Section~\ref{sec:Bluing} we explore possible explanations for this phenomenon, and in Section~\ref{sec:Discussion} we discuss the implications on the configuration of the system. Appendix~\ref{sec:ObsConditions} provides more details regarding the observing conditions during the ULTRACAM run. Appendix~\ref{sec:LC} presents the light curve evolution, as observed during our week-long observing campaign by a collection of telescopes around the globe, combined with the months-long light curve provided by \citet{Gary_2017}.

\section{ULTRACAM observations and data analysis}
\label{sec:Obs}

\begin{figure*}
\includegraphics[width=0.9\textwidth]{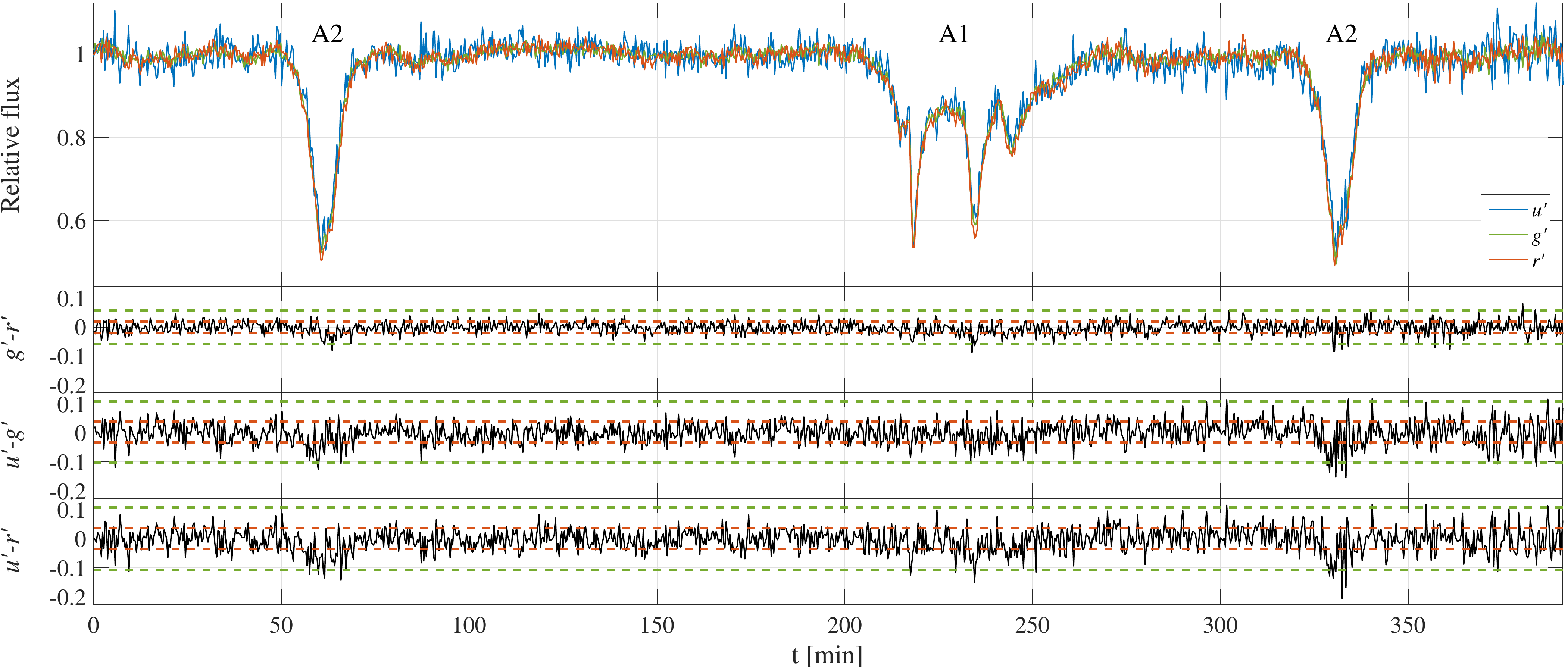}
\caption{Top: ULTRACAM light curve from 2016 April 21, in \ub- (blue), \gb- (green), and \rb-band (red), binned to an effective exposure time of 20\,s. Bottom: colour indices, as marked (black solid line). The dashed lines indicate the $1\sigma$ (red) and $3\sigma$ (green) levels. The labels above the dips denote dips which share a common period. Note the 2-3$\sigma$ detection of bluing during transits only, as evidenced by negative values of $\ub-\gb$ and $\ub-\rb$.}
\label{fig:Color20160421}
\end{figure*}

\begin{figure*}
\includegraphics[width=0.9\textwidth]{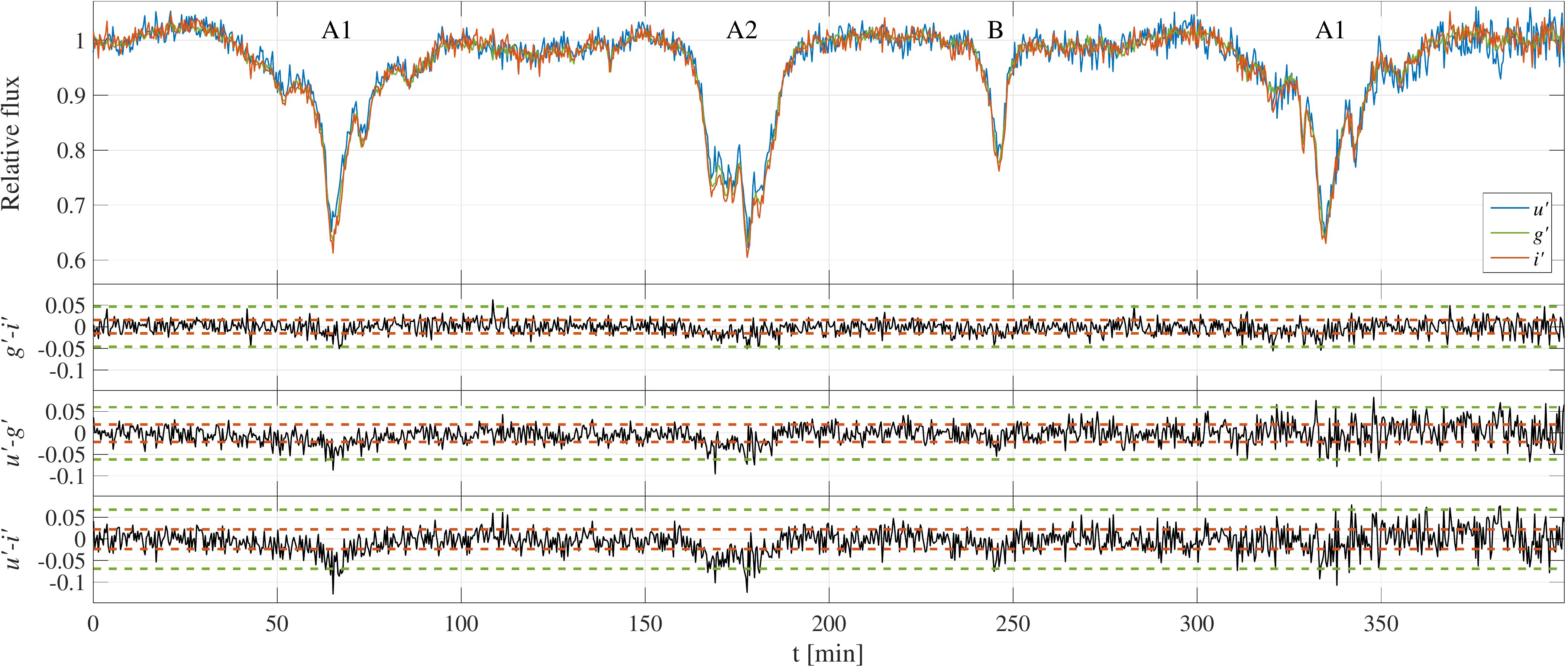}
\caption{Same as Fig.~\ref{fig:Color20160421}, for 2016 April 26, with \ib-band instead of \rb-band. Note the significant evolution of the light curve in merely five nights: the change in the shape of the `A1' and `A2' dips, and the appearance of the `B'-dip (see Appendix~\ref{sec:LC}).}
\label{fig:Color20160426}
\end{figure*}

We obtained multi-band fast photometry using ULTRACAM \citep{Dhillon_2007}, a visitor instrument mounted on the European Southern Observatory (ESO) 3.6\,m New Technology Telescope (NTT) at the La Silla Observatory, Chile, under Programme 097.C-0829 (PI: Hallakoun). Due to the weather conditions we were able to observe only on two of our six awarded nights (2016 April 21 and 26), covering almost 1.5 cycles ($\sim 6.7$\,h) each night. Although there were some passing clouds during the observations, they were mostly out-of-transit and did not have a significant effect on the shape of the light curve due to the relative nature of the observations (see Appendix~\ref{sec:ObsConditions}, which also describes the observational errors, for details). ULTRACAM is a high-speed camera capable of obtaining fast photometry of faint objects in three bands simultaneously with a negligible dead time ($\sim 25$\,ms) between exposures. We used SDSS \ub, \gb, and \rb\ filters on the first night, and \ub, \gb, and \ib\ filters on the second night. The CCD was windowed to achieve 5\,s exposure times in the slow readout mode (except in the \ub-band, where 10\,s exposures were obtained). The data were bias and flat-field corrected using the UTLRACAM pipeline \citep{Dhillon_2007}, which was then used to obtain aperture photometry of the target, using a nearby star as a reference. The light curve in each band was divided by a parabolic fit to its out-of-transit parts, to eliminate any systematics due to atmospheric extinction in the presence of colour differences between the WD and the reference star. Only the most featureless parts of the light curves were chosen as the out-of-transit intervals (see Figs~\ref{fig:Reduction1} and \ref{fig:Reduction2}). The same intervals were used in all bands. The aperture size was scaled according to the varying full width at half-maximum of the stellar profile. Figs~\ref{fig:Color20160421} and \ref{fig:Color20160426} show the reduced differential photometry from the two nights.

\begin{table*}
\caption{Mean transit depth of the dip-features that appear in the ULTRACAM light curves, in the \ub\gb\rb\ib-bands, calculated using equation~\ref{eq:MeanDepth}. The transit depth was averaged over the entire dip-group transit duration, $\Delta t$.}
\label{tab:MeanTransitDepth}
\begin{center}
\begin{tabular}{c c c c c c c c}
\hline
Dip & \multicolumn{2}{c}{Transit time (UT)} & $\Delta t$ [min]& $\overline{D}_{\ub}$ & $\overline{D}_{\gb}$ & $\overline{D}_{\rb}$ & $\overline{D}_{\ib}$ \\
\hline
\multirow{3}{*}{A1} & 2016-04-22 & 02:59:44-04:08:55 & 69.2 & $12.07\%\pm0.26\%$ & $12.277\%\pm0.091\%$ & $12.95\%\pm0.11\%$ & --- \\
 & 2016-04-26/27 & 23:49:59-00:50:47 & 60.8 & $8.81\%\pm0.17\%$ & $9.724\%\pm0.057\%$ & --- & $9.79\%\pm0.11\%$ \\
 & 2016-04-27 & 04:21:27-05:19:08 & 57.7 & $10.50\%\pm0.21\%$ & $9.936\%\pm0.069\%$ & --- & $10.38\%\pm0.12\%$ \\
\hline \\
\multirow{3}{*}{A2} & 2016-04-22 & 00:28:03-00:55:05 & 27.0 & $13.72\%\pm0.41\%$ & $14.77\%\pm0.14\%$ & $15.32\%\pm0.18\%$ & --- \\
 & 2016-04-22 & 04:58:25-05:20:54 & 22.5 & $17.93\%\pm0.48\%$ & $18.62\%\pm0.17\%$ & $18.90\%\pm0.21\%$ & --- \\
 & 2016-04-27 & 01:49:55-02:28:42 & 38.8 & $14.54\%\pm0.20\%$ & $15.221\%\pm0.069\%$ & --- & $15.90\%\pm0.12\%$ \\
\hline \\
\multirow{1}{*}{B} & 2016-04-27 & 03:14:25-03:28:04 & 13.6 & $9.14\%\pm0.38\%$ & $9.85\%\pm0.13\%$ & --- & $10.60\%\pm0.22\%$ \\
\hline \\

\end{tabular}
\end{center}
\end{table*}

Table~\ref{tab:MeanTransitDepth} lists the mean transit depths of the main features in the ULTRACAM light curves. The observed mean transit depth, $\overline{D}$, was measured by integrating over the in-transit light curve:
\begin{equation}
\label{eq:MeanDepth}
\overline{D} = 1 - \frac{1}{\Delta t}\sum_{i \in \textrm{ transit}}f_i dt_i
\end{equation}
where $\Delta t$ is the transit duration, $f$ is the normalised flux, and $dt$ is the sample interval. As seen in Figs~\ref{fig:Color20160421} and \ref{fig:Color20160426}, our ULTRACAM observations reveal a clear colour difference between the in- and out-of-transit photometry ($\ub-\rb \sim -0.05$\,mag). Although this is only a 2-3$\sigma$ detection, the fact that it occurred only during transits reinforces its significance. Surprisingly, the observed colour difference indicates `bluing', manifested by deeper transits in the redder bands, and not the usual reddening observed in dusty environments.

\section{Possible bluing mechanisms}
\label{sec:Bluing}

The cause of the blue flux excess during transit could be some property of the surface of the WD itself, some property of the obscuring material, or some other component or configuration of the circumstellar environment. We investigate below several possibilities.

\subsection{Bluing: limb darkening?}

\begin{figure}
\centering
\includegraphics[width=0.8\columnwidth]{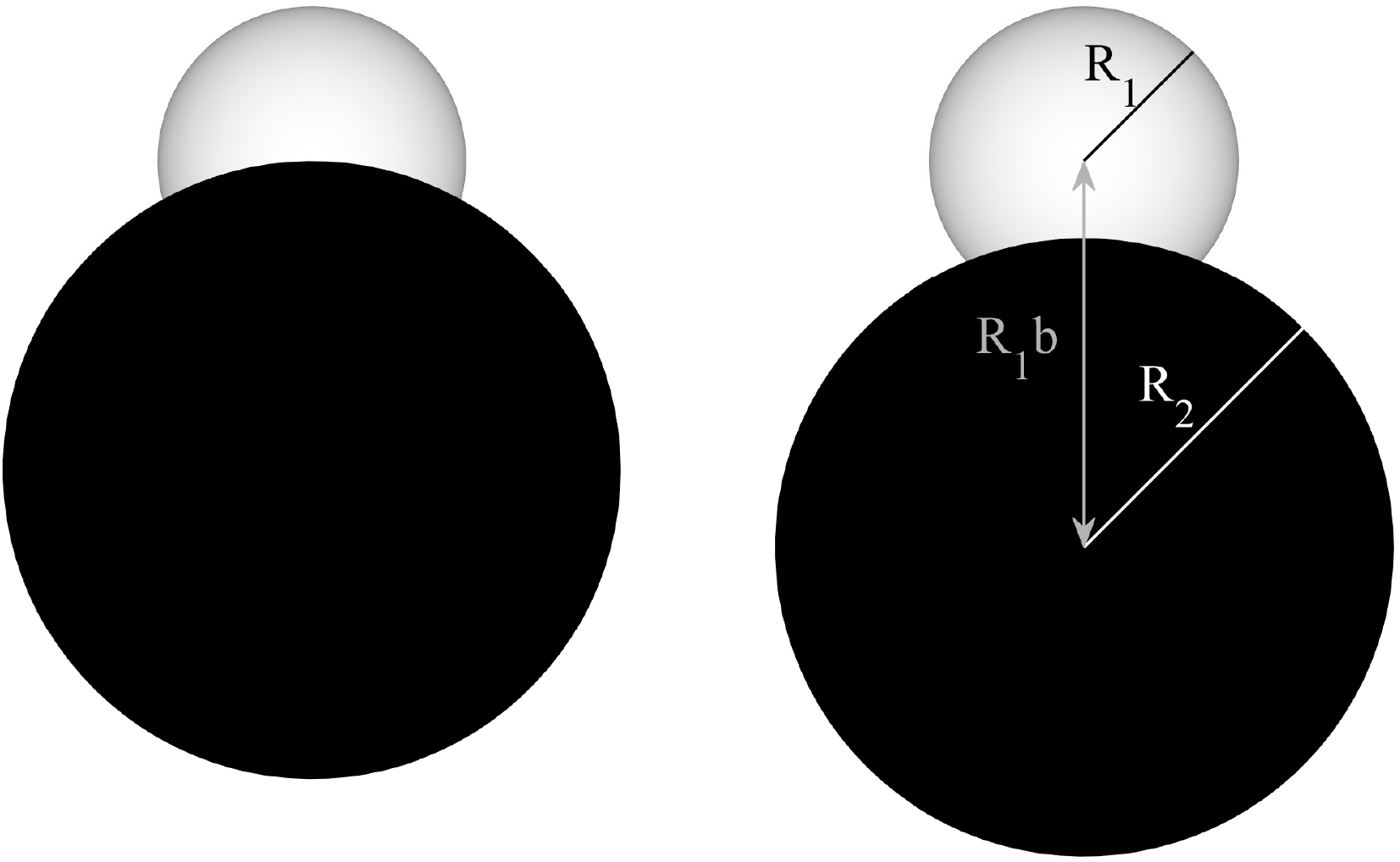}
\caption{Transit configurations, using $R_2/R_1=2$. Left: the minimal impact parameter required for bluing, $b = R_2/R_1$; right: the configuration yielding the maximal amount of bluing, $b = R_2/R_1 + 0.5$. The illustrations use the \ub-band limb-darkening profile.}
\label{fig:TransitConfig}
\end{figure}

\begin{figure}
\includegraphics[width=\columnwidth]{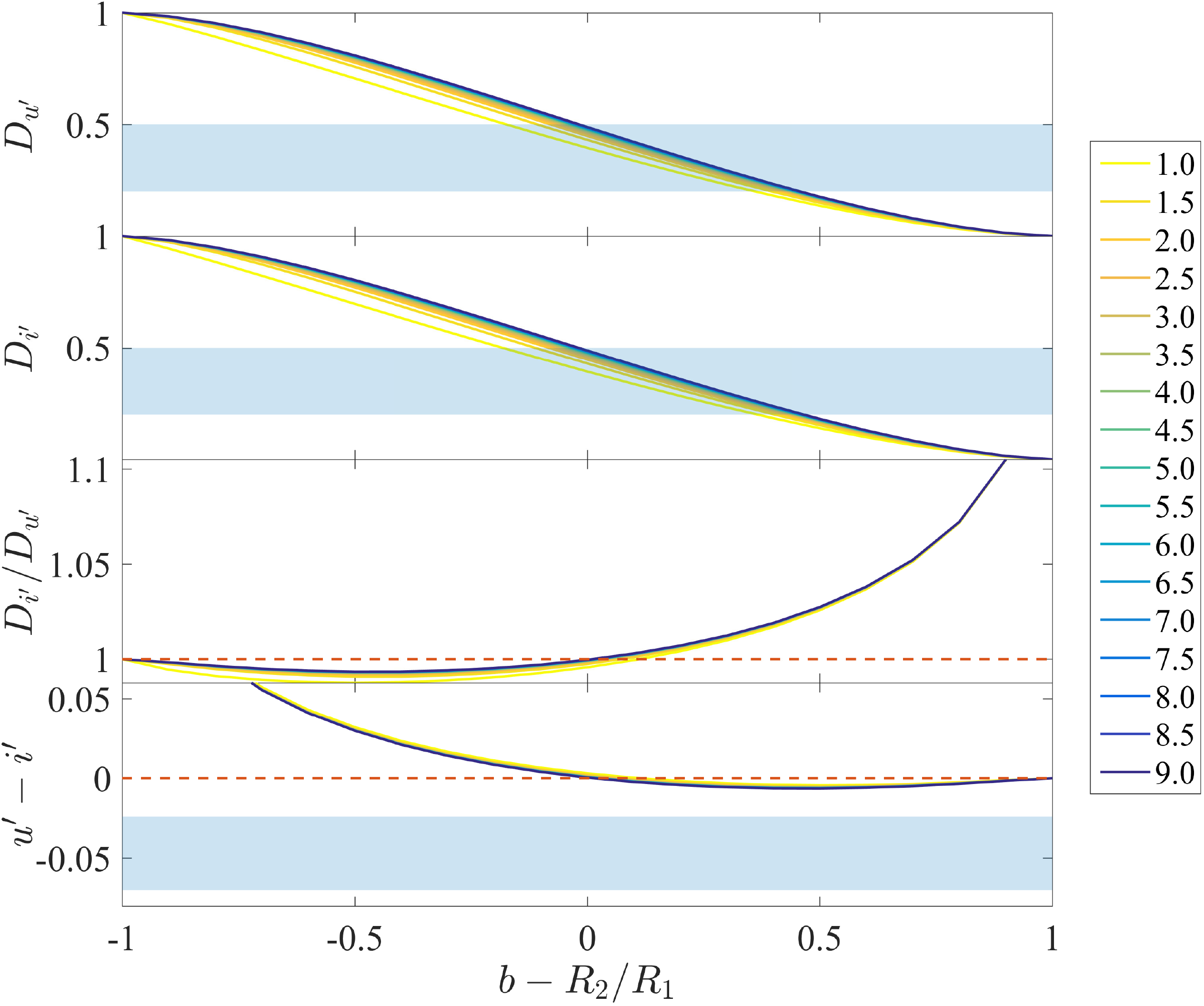}
\caption{Limb-darkening simulation. From top to bottom: \ub-band transit depth, \ib-band transit depth, \ib-band to \ub-band transit depth ratio, and \ub-\ib-band colour index, as a function of the projected distance of the WD centre from the edge of the occulting disc at the time of mid-transit. The solid lines corresponds to different $R_2/R_1$ values (see legend). The blue-shaded areas in the top two plots mark the approximate observed transit depth range (20-50~per cent), while in the bottom plot it marks the 1 to 3$\sigma$ range of the colour detection ($-0.06 \leq \ub-\ib \leq -0.02$\,mag). The dashed red lines mark the bluing threshold ($D_{\ib}/D_{\ub} > 1$, $\ub-\ib < 0$). The maximal amount of bluing produced by limb darkening is much smaller than observed, and its required transit configuration cannot explain the observed transit depths.}
\label{fig:LDsim}
\end{figure}

Since the limb of a WD is cooler and redder than its centre, a transiting object that obscures mainly the limb might explain the deeper transits observed in the redder bands. To test this assumption, we calculated the expected transit depths in the \ub- and \ib-band, over a grid of $R_2/R_1$ and $b$ values, where $R_2/R_1$ is the ratio of the radius of the occulting object (assumed, for simplicity, to be of circular cross section) to the radius of the WD, and $b = \left(a/R_1\right) \cos i$ is the impact parameter, where $i$ is the orbital plane's inclination to the line of sight, and $a$ is the orbital separation between the centres of the WD and the transiting object. We estimated the limb-darkening profiles for WD\,1145+017's effective temperature and surface gravity \citep[$T_\textrm{eff}=15900$\,K, $\log g = 8$, see][]{Vanderburg_2015} using Claret's limb-darkening coefficients for hydrogen-dominated DA-type WDs, as calculated by \citet{Gianninas_2013}. The coefficients used were $0.73$, $-0.53$, $0.67$, and $-0.27$ for the \ub-band, and $0.61$, $-0.40$, $0.34$, and $-0.12$ for the \ib-band. Fig.~\ref{fig:TransitConfig} illustrates two simulated configurations. Fig.~\ref{fig:LDsim} shows the calculated \ib- to \ub-band transit depth ratio, $D_{\ib}/D_{\ub}$, and the \ub-\ib-band colour index during transit, all as a function of the projected distance of the WD centre from the edge of the occulting disc, $b - R_2/R_1$, at the time of mid-transit. The WD was modelled as a disc divided into $1000$ concentric annuli, while the occulting object was modelled as a uniformly opaque disc. The results indicate a small amount of bluing for $b \gtrsim R_2/R_1$, with a maximum at $b = R_2/R_1 + 0.5$. The expected \ub-\ib-band colour index for the system configuration with the maximal amount of bluing ($\sim -0.006$\,mag), misses the observed amount of bluing ($\sim -0.05$\,mag) by an order of magnitude. Thus, while some bluing can be achieved with a grazing transit, covering only the limb of the WD, it is not sufficient. In addition, the $\sim 0.4$ transit depth requires the occulting object to cover a significant portion of the more central parts of the WD disc, in contrast to the grazing scenario. Replacing the opaque disc used to model the occulting object by a disc with varying opacity will further reduce the amount of bluing, aggravating the problem. We should note that the limb-darkening profiles used here are for DA-type WDs, while WD\,1145+017 is a metal-polluted helium-dominated DBAZ-type WD. Although the effect this difference might have on the limb-darkening profiles requires further investigation, it seems unlikely to provide a limb-darkening solution to the problem.

\subsection{Bluing: peculiar dust properties?}

Although bluing by dust is rare, it is possible. After the 1883 eruption of the Krakatoa volcano there were reports of a blue moon and a violet sun \citep{Bohren_1983}. Bluing by dust is common on Mars, where blue sunsets have been captured by NASA's Curiosity rover camera\footnote{\url{http://www.jpl.nasa.gov/spaceimages/details.php?id=pia19400}}. Bluing by dust is possible only for specific and narrow distributions of grain sizes, such as the one observed on Mars \citep{Fedorova_2014}.

Following \citet{Hansen_1971} and \citet{Hansen_1974}, we define the effective radius of a distribution of dust grains, $r_\textrm{eff}$, which is the mean grain radius weighted by its geometrical cross section,
\begin{equation}
r_\textrm{eff} = \frac{\int_0^\infty r \pi r^2 n\left(r\right) dr}{\int_0^\infty \pi r^2 n\left(r\right) dr},
\end{equation}
and the dimensionless effective variance, $v_\textrm{eff}$,
\begin{equation}
v_\textrm{eff} = \frac{\int_0^\infty \left(r - r_\textrm{eff}\right)^2 \pi r^2 n\left(r\right) dr}{r_\textrm{eff}^2 \int_0^\infty \pi r^2 n\left(r\right) dr},
\end{equation}
where $n \left(r\right)$ is the grain-size distribution. Grain-size distributions sharing the same effective radius and variance display the same scattering characteristics \citep{Hansen_1971}. We use the `standard' grain size distribution of \citet{Hansen_1971},
\begin{equation}
n \left( r \right) \propto r^{\frac{\left(1-3v_\textrm{eff}\right)}{v_\textrm{eff}}} e^{-\frac{r}{r_\textrm{eff}v_\textrm{eff}}},
\end{equation}
which is characterised by the effective radius and variance.
The mean extinction cross-section of the grain-size distribution is then \citep[][eq. 3]{Croll_2014}:
\begin{equation}
\overline{\sigma}_\textrm{ext} \left(r_\textrm{min}, r_\textrm{max}, \lambda \right) = \frac{\int_{r_\textrm{min}}^{r_\textrm{max}} n\left(r'\right) \sigma_\textrm{ext}\left(r', \lambda\right)dr'}{\int_{r_\textrm{min}}^{r_\textrm{max}} n\left(r'\right)dr'},
\end{equation}
where $\sigma_\textrm{ext}\left(r', \lambda\right)$ is the extinction cross-section as a function of the grain size and the wavelength.

\begin{figure}
\includegraphics[width=\columnwidth]{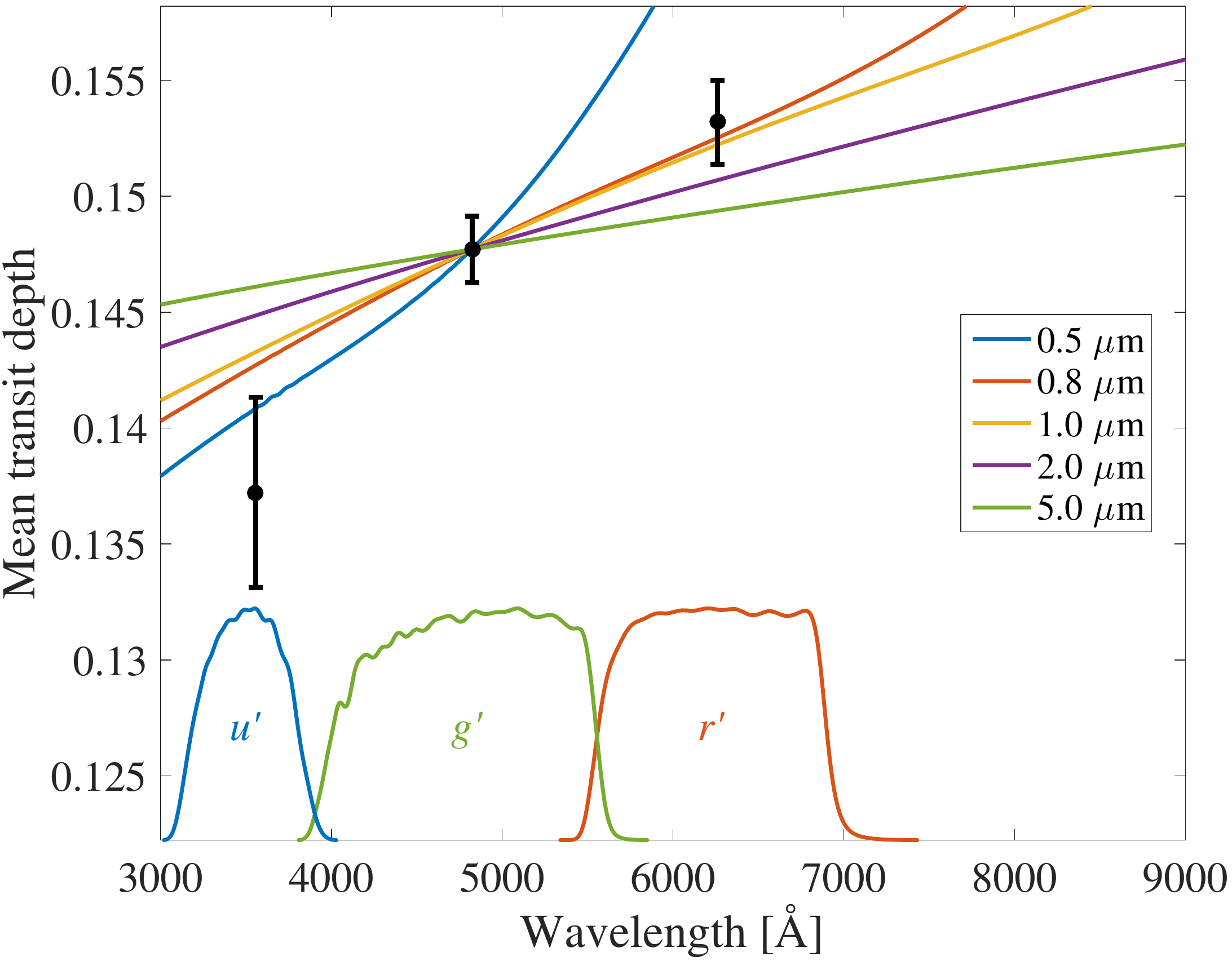}
\caption{The mean transit depth in \ub-, \gb-, and \rb-bands (black errror bars) for the first `A2' dip in the ULTRACAM light curve from 2016 April 21. The filter bandpasses are shown below each errorbar. The coloured curves correspond to the extinction cross-section curves of `astronomical silicate' for various effective grain radii (see legend), all scaled to fit the \gb-band measurement. The curves for $\sim 0.8-2\,\mu$m-sized grains can match the observed bluing, but rather marginally.}
\label{fig:Extinction1}
\end{figure}

\begin{figure}
\includegraphics[width=\columnwidth]{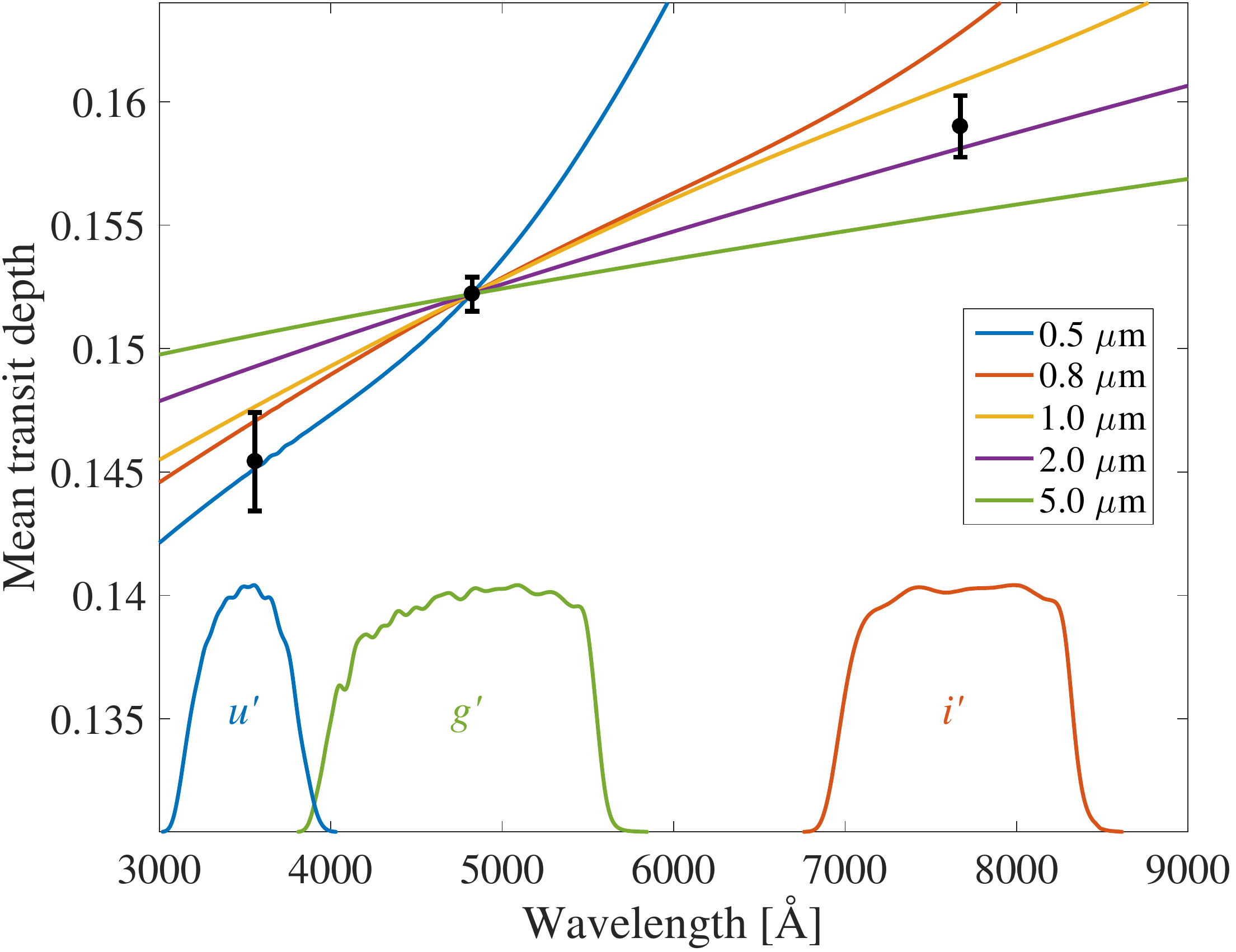}
\caption{Same as Fig~\ref{fig:Extinction1}, for the `A2' dip in the ULTRACAM night from 2016 April 26, and with \ib-band instead of \rb-band.}
\label{fig:Extinction2}
\end{figure}

We calculated the mean extinction cross-section as a function of wavelength for grain-size distributions of effective radii ranging between 0.5 and 10\,$\mu$m, with effective variance of 0.1. The extinction cross-sections were calculated using a \textsc{Matlab} implementation of the Mie scattering code of \citet[][appendix A]{Bohren_1983}. We used the `astronomical silicate' refractive index of \citet{Draine_2003}\footnote{See \url{https://www.astro.princeton.edu/~draine/dust/dust.diel.html} and \url{ftp://ftp.astro.princeton.edu/draine/dust/diel/callindex.out_silD03}.}. In the wavelength range $\lambda = 3000-9000$\,$\AA$, $n$, the real component of this refractive index, varies between 1.69 and 1.73, while the imaginary part, $k$, varies between 0.029 and 0.031. The effective radius serves as a good characterising parameter for the grain-size distribution, as long as $\pi r_\textrm{eff} \left(n - 1\right) \gtrsim \lambda$ \citep{Hansen_1971}, i.e. for $r_\textrm{eff} \gtrsim 0.4\,\mu$m. Figs~\ref{fig:Extinction1} and \ref{fig:Extinction2} compare the mean transit depth as measured in the various bandpasses for the two `A2' transits on two nights (see Figs~\ref{fig:Color20160421} and \ref{fig:Color20160426}, and Table~\ref{tab:MeanTransitDepth}), with extinction cross-section curves for various effective radii. The extinction curves were scaled to match the \gb-band measurement. Different appearances of different dip-groups have different mean transit depth slopes. The two `A2' dips chosen here both have a significant transit depth (hence a good signal-to-noise ratio) and were measured at relatively low airmasses (minimizing possible systematics due to atmospheric extinction; see Figs~\ref{fig:Reduction1} and \ref{fig:Reduction2}).

We see that the observed level of inter-band bluing can, in principle, be achieved by invoking particular narrow grain-size distributions, such as the calculated $\sim 0.8-2\,\mu$m-grain-size extinction curves. These curves can simultaneously match the mean transit depth measured in all bands, but with a rather poor fit that deviates at the $1-2\sigma$ level. Other possible compositions have refractive indices similar to that of the silicate, except iron \citep{Croll_2014}. Replacing the `astronomical silicate' dust with a pure iron composition \citep[$1.67\leq n \leq 2.96$, $1.99 \leq k \leq 3.58$, see][]{Johnson_1974}\footnote{M. N. Polyanskiy, ``Refractive index database'', \url{http://refractiveindex.info}} results in an even worse match. Thus, although they cannot be completely ruled out, peculiar dust properties appear unlikely to be the explanation for the observed bluing during transits in WD\,1145+017, especially considering that at least some of the bluing must be the result of circumstellar absorption lines, as detailed below. A wider wavelength coverage, that includes the IR, might better constrain the particle size. 

\subsection{Bluing: circumstellar lines?}
\label{sec:CS}

\begin{figure*}
\includegraphics[width=\textwidth]{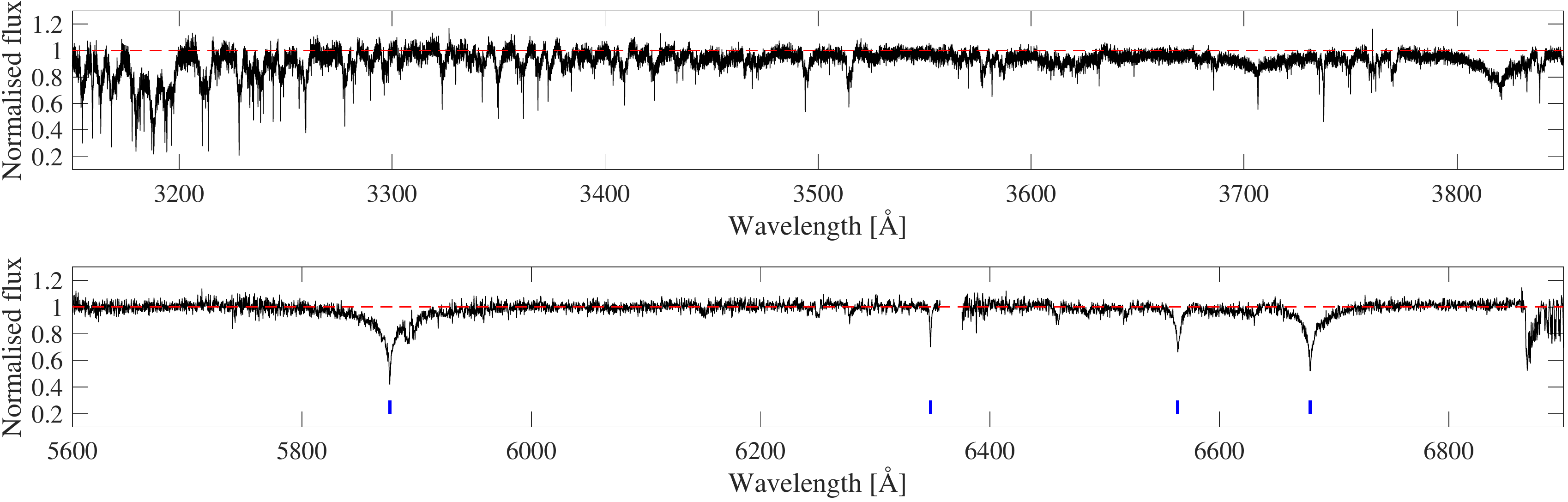}
\caption{Top: Normalised averaged Keck/HIRES spectrum from 2015 April 11 over the \ub-band range \citep{Xu_2016}. Note the increasing density of circumstellar absorption lines toward the UV side of the spectrum. Bottom: Normalised averaged VLT/X-SHOOTER spectrum from 2016 March 29, over the \rb-band range. The blue ticks mark the photospheric absorption lines (other features are telluric).}
\label{fig:Spectrum}
\end{figure*}

WD\,1145+017 displays unique broad absorption lines induced by circumstellar gas \citep{Xu_2016}. When observed spectroscopically during transits, these lines occasionally appear shallower \citep[][Xu et al., in preparation]{Redfield_2016}. Interestingly, most of these lines populate the \ub-band range ($\sim 3257-3857\,\AA$, see Fig.~\ref{fig:Spectrum}). Hence, the \ub-band excess during transits might be explained by the reduced circumstellar absorption along the line-of-sight.

\begin{figure}
\includegraphics[width=\columnwidth]{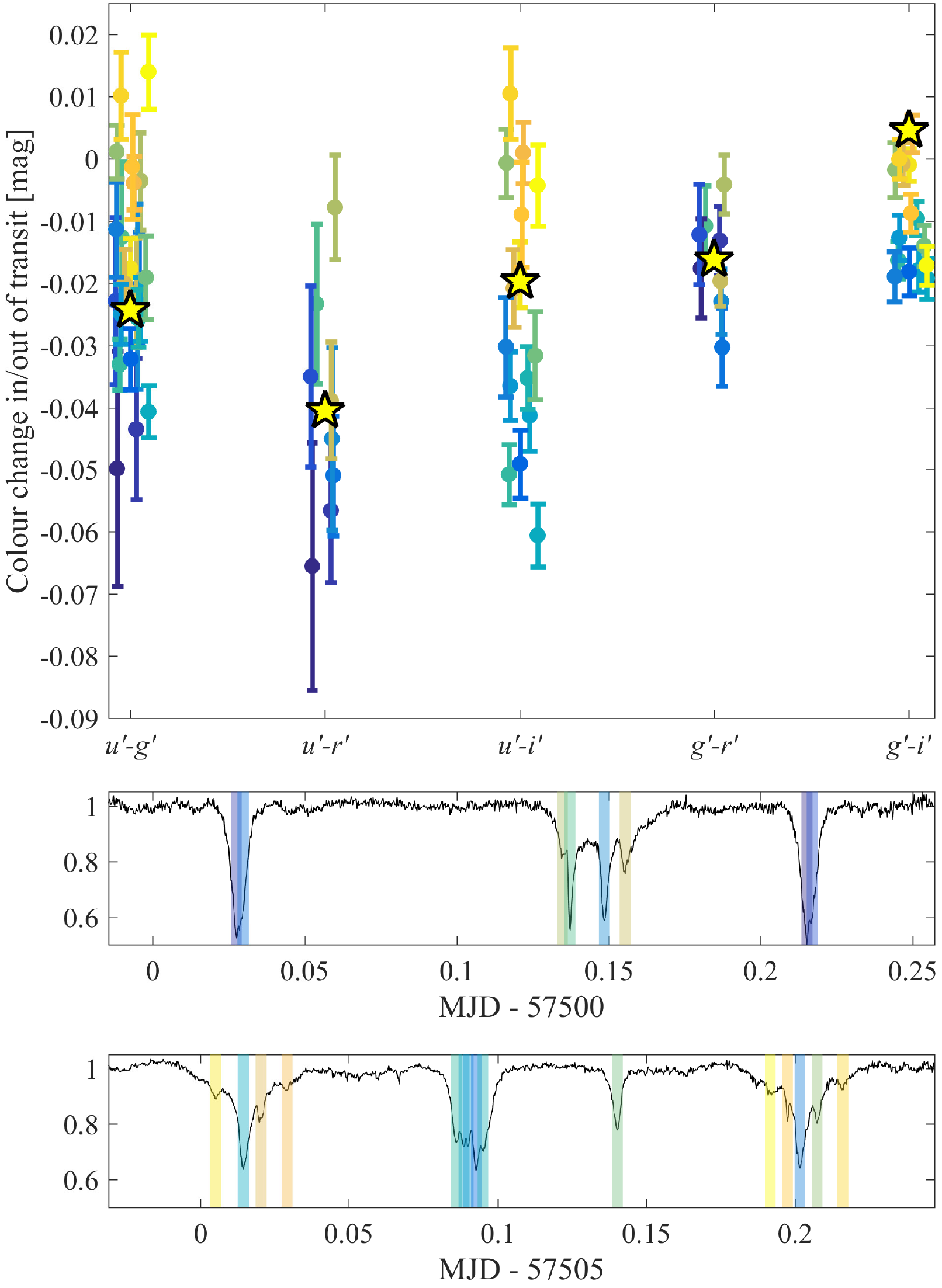}
\caption{Demonstration of the circumstellar gas-induced bluing using VLT/X-SHOOTER spectra: Photometric (error bars) and `spectral' (star-shape symbols) colour differences in- and out-of-transit, for various colour indices. The `spectral' colour measurements were performed on X-SHOOTER spectra obtained on 2016 March 29 (see Section~\ref{sec:CS} for details). The error bars are colour-coded by the corresponding mean transit depth (the bluer the deeper). The lower panels show the \gb-band ULTRACAM light curves from 2016 April 21 (middle) and 2016 April 26 (bottom), binned to an effective exposure time of 20\,s. The highlighted areas correspond to the integrated time intervals, with colours matching those of the top panel error bars. The photometry derived from the spectra broadly reproduces the bluing and the trends seen in the photometric transit data.}
\label{fig:ColorComparison}
\end{figure}

To test this assumption, we have used in- and out-of-transit spectra, integrated over the various photometric bandpasses, to calculate `spectral' photometry measurements in the \ub-, \gb-, \rb-, and \ib-bands. The spectra were obtained using X-SHOOTER on the ESO Very Large Telescope (VLT) on 2016 March 29, under Director's Discretionary Time Programmes 296.C-5024 (PI: Xu) and 296.C-5014 (PI: Farihi; see \citealt{Redfield_2016} for details). The wide wavelength range of X-SHOOTER (3000 to 10,000\,$\AA$) provides simultaneous coverage of the \ub-\ib-bands in a single exposure. The out-of-transit spectrum used here was taken at 03:37~(UT), while the in-transit spectrum is from 04:37~(UT), both with 280\,s and 314\,s exposures in the UVB and VIS arms, respectively. The spectra were smoothed using 5-point span robust local regression (RLOWESS) to get rid of spurious outlier points resulting from imperfect sky subtraction. Since we are interested in the spectral changes resulting solely from the circumstellar lines, the in-transit spectrum was scaled to fit the out-of-transit spectrum's continuum level. Finally, both spectra were integrated over the various ULTRACAM bandpasses and the colours were calculated (indicated by the star-shape symbols in Fig.~\ref{fig:ColorComparison}).

For a comparison, we integrated the various colour measurements of the ULTRACAM light curves over 314\,s intervals centred around the main transit features. From the light curves published by \citet[][see Appendix~\ref{sec:LC} below]{Gary_2017}, we can associate the transit detected in the X-SHOOTER data with the `A1' dips seen in the ULTRACAM light curves (see Appendix~\ref{sec:LC}). However, as is evident from the significant changes that this group of dips undergoes within a few hours, we do not expect that the transit detected by X-SHOOTER about a month earlier will match exactly our April observations. A deeper transit implies a larger obscured area, which is likely to induce a more significant colour change. This is hinted at in Fig.~\ref{fig:ColorComparison}, where there appears to be a correlation between the mean transit depth (indicated by the colour-code) and the observed colour difference.

Taking all of these considerations into account, there seems to be a relatively good agreement between the photometric and `spectral' colour differences. Moreover, repeating the spectral photometry calculation for two out-of-transit epochs (as well as two in-transit epochs) revealed no significant colour difference between the epochs, as expected.

To investigate more directly whether the change in absorption depth is responsible for the bluing seen in the spectral photometry, we have estimated whether the total absorption equivalent width (EW) over the \ub-band range is indeed significant. Fig.~\ref{fig:Spectrum} (top) shows the \ub-band range of the WD\,1145+017 spectrum taken by \citet{Xu_2016} using the HIgh-Resolution Echelle Spectrometer (HIRES) on the Keck 10\,m telescope \citep{Vogt_1994}. This spectrum is the average of three 2400\,s exposures taken on 2015 April 11 (see \citealt{Xu_2016} for further details). The spectrum has been divided by a parabolic fit to line-free regions in the redder parts of the \ub-band in the spectrum. The total \ub-bandpass-weighted absorption equivalent width (EW) is $\gtrsim 34\,\AA$, with a relative absorption of $\gtrsim 8$~per cent over the \ub-bandpass. This is a lower limit on the EW, as it is clear that, as one approaches the ultraviolet, line blanketing sets in, the absorptions overlap, and the true continuum level could be significantly above the observed flux levels. The true total EW could easily be a factor 2 higher. However, this possibility cannot be tested directly using the HIRES spectra because the wavelength response calibration of the spectra is not accurate to the few-percent levels required for such a comparison. Furthermore, the activity level of the system has increased significantly during the year that has passed between the HIRES observations of \citet{Xu_2016} and our ULTRACAM run, both in terms of transit depth \citep[see fig. 10 of][]{Gary_2017} and circumstellar line strength \citep[][Xu et al., in preparation]{Redfield_2016}. Indeed, a similar Keck/HIRES spectrum from April 2016 (Xu et al., in preparation) shows that the total \ub-bandpass-weighted EW has grown to at least 50\,$\AA$, with a relative absorption of $\gtrsim 11$~per cent.

In addition, a preliminary analysis of line strength variations in- and out-of-transit of a single, strong, visual-band circumstellar absorption line, \ion{Fe}{ii} 5169\,$\AA$, shows that during the deepest transits, the line depth can decrease by up to $\sim 30$~per cent (Xu et al., in preparation). If the relative absorption by circumstellar lines in the \ub-band is, say, 12~per cent, then a 30~per cent reduction in EW during transit could yield the observed 0.04\,mag change in colour between the \ub-band and some other band (e.g. the \rb-band, see Fig.~\ref{fig:Spectrum}, bottom) that has relatively few absorptions.

The above tests argue that reduced circumstellar absorption during transits is the most plausible explanation for the observed bluing. The fact that previous studies \citep{Croll_2017, Alonso_2016, Zhou_2016} did not observe wavelengths shorter than $4800\,\AA$, could explain why they did not detect bluing. However, this conclusion is not definitive because of the various complications: the difficulty in establishing the continuum level in the \ub-band and hence the total absorption EW, and the variance of the line strength change during transits. Future colour measurements in bands having a high density of circumstellar absorption lines should take this into account. Simultaneous photometric and spectroscopic measurements would provide a clearer understanding of the bluing mechanism.

\section{Discussion}
\label{sec:Discussion}
Assuming that the bluing that we have observed during transits in WD\,1145+017 is the result of a reduced absorption by the circumstellar gas during transit, we can constrain the possible configurations of the components of the system. First, the fact that the circumstellar absorption is affected by the transits implies that at least part of the absorbing gas shares the same plane as the transiting objects. Second, we can constrain the location of the gas relative to the WD and the transiting objects. Based on the observed bluing alone, without taking into account further constraints (see below), the circumstellar absorbing gas could be between the WD and the transiting objects, in front of both the WD and the debris, or a combination of the two (e.g. if the WD and the debris are embedded in the gas).

In the case that the gas is between the transiting objects and the WD, the gas cannot cover the WD uniformly; the reduction in the relative absorption depth during transit means that the debris is occulting more of the WD's line-absorbed light than the unabsorbed WD continuum emission (e.g. as in the geometry of a flat disc around the WD). Alternatively, the gas could be outside the debris' orbit, i.e. in front of both the debris and the WD. If the absorption lines are in the linear regime of their curve of growth, then no change in relative absorption depth would be expected unless, again, the gas does not cover the WD uniformly. In the third, `embedded', possibility, at least some of the gas, whether in front or between the WD and the debris, must again cover the WD non-uniformly. One configuration satisfying all of these constraints is a circumstellar gas disc, coplanar with the debris orbit. The transits, the circumstellar lines, and the NIR excess, could be manifestations of the same general disc structure consisting of gas, dust, and planetesimal debris, although not necessarily all at the same radii.

If we take into account the spectroscopic and NIR observations as well, we can further constrain the possible configurations of the system. The out-of-transit depth of the circumstellar absorption lines is consistent with the non-uniform coverage constraint. The significant broadening of the circumstellar lines \citep[][Xu et al., in preparation]{Xu_2016, Redfield_2016} could be the result of the line-of-sight velocities of gas in a mildly-eccentric edge-on disc, in a close-in orbit between the WD and the transiting objects \citep{Redfield_2016}.
\citet{Rappaport_2016} have suggested that the fragments causing the transits come from an asteroid that is filling its critical potential surface at a relatively stable orbit around the `A'-period. When the asteroid passes near its L1 point, fragments occasionally break off and drift into a slightly inner orbit. The dust disc is not well-constrained by the observations so far. \citet{Vanderburg_2015} argued that the detected amount of NIR excess indicates that an optically-thick dust disc could not be observed edge-on (i.e. it is misaligned with the transiting objects). However, \citet{Gary_2017} counter that an edge-on optically-thick dust disc could yield the observed amount of NIR excess if it extends from the transiting objects' `A'-period orbit ($\sim 94\,R_\textrm{WD}$) to about $140\,R_\textrm{WD}$. \citet{Zhou_2016} argue for an edge-on optically-thin dust ring, located within the transiting objects' orbital range ($\sim 90-100\,R_\textrm{WD}$), based on the lack of optical reddening in the WD's spectral energy distribution (SED). Our proposed configuration is consistent with such an edge-on dust ring scenario.

\section{Conclusions}
Using simultaneous multi-band fast photometry of the debris transits around WD\,1145+017, we have detected `bluing' -- deeper transits in the redder bands -- and not reddening, as commonly detected in dusty environments. This bluing cannot be the result of limb darkening, since there seems to be no configuration that simultaneously yields the observed transit depth and the required amount of bluing. Bluing as a result of peculiar dust properties is a possible explanation, but marginally so, as it seems hard to find the fine-tuned grain-size distribution that will simultaneously and precisely reproduce the bluing observed between the various bands. The fact that most of the circumstellar absorption lines, which can be shallower during transits, are in the \ub-band, appears to provide a plausible explanation for the excess blue flux. The bluing by the circumstellar lines, which can explain most or all of the observed photometric effect, furthermore means that little or none of the effect remains to be explained by peculiar dust or, in other words, peculiar dust would predict bluing even larger than observed. In the circumstellar gas explanation, the requirements that the gas shares the debris orbit's inclination and that the gas only partly covers the lines of sight to the WD surface, both suggest that the circumstellar gas is in a disc. This could be, in principle, the same disc in which resides the debris (which produces the transits) or the dust (which produces a NIR excess). However, it is more likely to be a distinct disc, at a smaller orbital radius, based on the broadening of the circumstellar lines.

Our results indicate that the circumstellar gas affects the photometric light curve, at least to some degree. Future simultaneous photometric and spectroscopic observations of the system would clear up much of the uncertainty regarding the interpretation of the bluing.

The combined light curve of WD\,1145+017, using our week-long monitoring campaign and the months-long light curve of \citet{Gary_2017}, reveals several intriguing features \citep[most of them already mentioned by][]{Gary_2017}, described in Appendix~\ref{sec:LC}. These include the breaking up of the `A2' group, the first detection of the `B'-period, and the detection of smaller transits in the high signal-to-noise ULTRACAM light curve. Our light curves are available for further study upon request, in the hope that they will help to constrain future dynamical models of the system.

\section*{Acknowledgements}

We are indebted to Bruce L. Gary, Saul Rappaport, Tom Kaye, Roi Alonso, and Josch Hambsch for providing their WD\,1145+017 light curves and dip fit results.
We thank the PIs, Michael Jura (U019), John Debes (U124D), and Ben Zuckerman (U067E), for providing the Keck spectra of WD\,1145+017 and sharing the results of the preliminary analysis. 
We thank the anonymous referee for valuable comments that improved this paper.
We thank Bruce L. Gary, Saul Rappaport, Ferdinando Patat, Tsevi Mazeh, Alexandros Gianninas, Sahar Shahaf, Yakov Faerman, Stefan Jordan, and Andreas Quirrenbach for useful discussions, the members of the VBO Steering Committee and the Time Allocation Committee (TAC), and the observing assistants: Mr G.~Selvakumar, Mr S.~Venkatesh, Mr S.~Parthiban and Mr R.~Rajini Rao.
NH thanks Tamar Faran for moral support.
This work was supported in part by Grant 1829/12 of the Israeli Centers for Research Excellence (I-CORE) programme of the Planning and Budgeting Committee (PBC) and the Israel Science Foundation (ISF) and by Grant 648/12 by the ISF (DM).
The ULTRACAM team acknowledges the support of the Science and Technology Facilities Council (STFC).
Based on observations made with ESO Telescopes at the La Silla Paranal Observatory under programme IDs 097.C-0829, 296.C-5024 and 296.C-5014.
This work makes use of observations from the LCO global telescope network.
UKIRT is supported by NASA and operated under an agreement among the University of Hawaii, the University of Arizona, and Lockheed Martin Advanced Technology Center; operations are enabled through the cooperation of the East Asian Observatory.
The UC Observatory Santa Martina was part of the photometric monitoring campaign presented in this work.
This work used the astronomy \& astrophysics package for Matlab (Ofek 2014).




\bibliographystyle{mnras}
\bibliography{WD1145}

\begin{thebibliography}{}
\makeatletter
\relax
\def\mn@urlcharsother{\let\do\@makeother \do\$\do\&\do\#\do\^\do\_\do\%\do\~}
\def\mn@doi{\begingroup\mn@urlcharsother \@ifnextchar [ {\mn@doi@}
  {\mn@doi@[]}}
\def\mn@doi@[#1]#2{\def\@tempa{#1}\ifx\@tempa\@empty \href
  {http://dx.doi.org/#2} {doi:#2}\else \href {http://dx.doi.org/#2} {#1}\fi
  \endgroup}
\def\mn@eprint#1#2{\mn@eprint@#1:#2::\@nil}
\def\mn@eprint@arXiv#1{\href {http://arxiv.org/abs/#1} {{\tt arXiv:#1}}}
\def\mn@eprint@dblp#1{\href {http://dblp.uni-trier.de/rec/bibtex/#1.xml}
  {dblp:#1}}
\def\mn@eprint@#1:#2:#3:#4\@nil{\def\@tempa {#1}\def\@tempb {#2}\def\@tempc
  {#3}\ifx \@tempc \@empty \let \@tempc \@tempb \let \@tempb \@tempa \fi \ifx
  \@tempb \@empty \def\@tempb {arXiv}\fi \@ifundefined
  {mn@eprint@\@tempb}{\@tempb:\@tempc}{\expandafter \expandafter \csname
  mn@eprint@\@tempb\endcsname \expandafter{\@tempc}}}

\bibitem[\protect\citeauthoryear{{Alonso}, {Rappaport}, {Deeg}  \&
  {Palle}}{{Alonso} et~al.}{2016}]{Alonso_2016}
{Alonso} R.,  {Rappaport} S.,  {Deeg} H.~J.,   {Palle} E.,  2016, \mn@doi
  [\aap] {10.1051/0004-6361/201628511}, \href
  {http://adsabs.harvard.edu/abs/2016A%26A...589L...6A} {589, L6}

\bibitem[\protect\citeauthoryear{{Althaus}, {C{\'o}rsico}, {Isern}  \&
  {Garc{\'{\i}}a-Berro}}{{Althaus} et~al.}{2010}]{Althaus_2010}
{Althaus} L.~G.,  {C{\'o}rsico} A.~H.,  {Isern} J.,   {Garc{\'{\i}}a-Berro} E.,
   2010, \mn@doi [\aapr] {10.1007/s00159-010-0033-1}, \href
  {http://adsabs.harvard.edu/abs/2010A%26ARv..18..471A} {18, 471}

\bibitem[\protect\citeauthoryear{{Bohren} \& {Huffman}}{{Bohren} \&
  {Huffman}}{1983}]{Bohren_1983}
{Bohren} C.~F.,  {Huffman} D.~R.,  1983, {Absorption and scattering of light by
  small particles}

\bibitem[\protect\citeauthoryear{{Brogi}, {Keller}, {de Juan Ovelar},
  {Kenworthy}, {de Kok}, {Min}  \& {Snellen}}{{Brogi}
  et~al.}{2012}]{Brogi_2012}
{Brogi} M.,  {Keller} C.~U.,  {de Juan Ovelar} M.,  {Kenworthy} M.~A.,  {de
  Kok} R.~J.,  {Min} M.,   {Snellen} I.~A.~G.,  2012, \mn@doi [\aap]
  {10.1051/0004-6361/201219762}, \href
  {http://adsabs.harvard.edu/abs/2012A%26A...545L...5B} {545, L5}

\bibitem[\protect\citeauthoryear{{Croll} et~al.,}{{Croll}
  et~al.}{2014}]{Croll_2014}
{Croll} B.,  et~al., 2014, \mn@doi [\apj] {10.1088/0004-637X/786/2/100}, \href
  {http://adsabs.harvard.edu/abs/2014ApJ...786..100C} {786, 100}

\bibitem[\protect\citeauthoryear{{Croll} et~al.,}{{Croll}
  et~al.}{2017}]{Croll_2017}
{Croll} B.,  et~al., 2017, \mn@doi [\apj] {10.3847/1538-4357/836/1/82}, \href
  {http://adsabs.harvard.edu/abs/2017ApJ...836...82C} {836, 82}

\bibitem[\protect\citeauthoryear{{Dhillon} et~al.,}{{Dhillon}
  et~al.}{2007}]{Dhillon_2007}
{Dhillon} V.~S.,  et~al., 2007, \mn@doi [\mnras]
  {10.1111/j.1365-2966.2007.11881.x}, \href
  {http://ukads.nottingham.ac.uk/abs/2007MNRAS.378..825D} {378, 825}

\bibitem[\protect\citeauthoryear{{Draine}}{{Draine}}{2003}]{Draine_2003}
{Draine} B.~T.,  2003, \mn@doi [\apj] {10.1086/379123}, \href
  {http://adsabs.harvard.edu/abs/2003ApJ...598.1026D} {598, 1026}

\bibitem[\protect\citeauthoryear{{Eastman}, {Siverd}  \& {Gaudi}}{{Eastman}
  et~al.}{2010}]{Eastman_2010}
{Eastman} J.,  {Siverd} R.,   {Gaudi} B.~S.,  2010, \mn@doi [\pasp]
  {10.1086/655938}, \href {http://adsabs.harvard.edu/abs/2010PASP..122..935E}
  {122, 935}

\bibitem[\protect\citeauthoryear{{Fedorova}, {Montmessin}, {Rodin}, {Korablev},
  {M{\"a}{\"a}tt{\"a}nen}, {Maltagliati}  \& {Bertaux}}{{Fedorova}
  et~al.}{2014}]{Fedorova_2014}
{Fedorova} A.~A.,  {Montmessin} F.,  {Rodin} A.~V.,  {Korablev} O.~I.,
  {M{\"a}{\"a}tt{\"a}nen} A.,  {Maltagliati} L.,   {Bertaux} J.-L.,  2014,
  \mn@doi [\icarus] {10.1016/j.icarus.2013.12.015}, \href
  {http://adsabs.harvard.edu/abs/2014Icar..231..239F} {231, 239}

\bibitem[\protect\citeauthoryear{{G{\"a}nsicke} et~al.,}{{G{\"a}nsicke}
  et~al.}{2016}]{Gaensicke_2016}
{G{\"a}nsicke} B.~T.,  et~al., 2016, \mn@doi [\apjl]
  {10.3847/2041-8205/818/1/L7}, \href
  {http://adsabs.harvard.edu/abs/2016ApJ...818L...7G} {818, L7}

\bibitem[\protect\citeauthoryear{{Gary}, {Rappaport}, {Kaye}, {Alonso}  \&
  {Hambschs}}{{Gary} et~al.}{2017}]{Gary_2017}
{Gary} B.~L.,  {Rappaport} S.,  {Kaye} T.~G.,  {Alonso} R.,   {Hambschs} F.-J.,
   2017, \mn@doi [\mnras] {10.1093/mnras/stw2921}, \href
  {http://adsabs.harvard.edu/abs/2017MNRAS.465.3267G} {465, 3267}

\bibitem[\protect\citeauthoryear{{Gianninas}, {Strickland}, {Kilic}  \&
  {Bergeron}}{{Gianninas} et~al.}{2013}]{Gianninas_2013}
{Gianninas} A.,  {Strickland} B.~D.,  {Kilic} M.,   {Bergeron} P.,  2013,
  \mn@doi [\apj] {10.1088/0004-637X/766/1/3}, \href
  {http://adsabs.harvard.edu/abs/2013ApJ...766....3G} {766, 3}

\bibitem[\protect\citeauthoryear{{Hansen}}{{Hansen}}{1971}]{Hansen_1971}
{Hansen} J.~E.,  1971, \mn@doi [Journal of Atmospheric Sciences]
  {10.1175/1520-0469(1971)028<1400:MSOPLI>2.0.CO;2}, \href
  {http://adsabs.harvard.edu/abs/1971JAtS...28.1400H} {28, 1400}

\bibitem[\protect\citeauthoryear{{Hansen} \& {Travis}}{{Hansen} \&
  {Travis}}{1974}]{Hansen_1974}
{Hansen} J.~E.,  {Travis} L.~D.,  1974, \mn@doi [\ssr] {10.1007/BF00168069},
  \href {http://adsabs.harvard.edu/abs/1974SSRv...16..527H} {16, 527}

\bibitem[\protect\citeauthoryear{{Johnson} \& {Christy}}{{Johnson} \&
  {Christy}}{1974}]{Johnson_1974}
{Johnson} P.~B.,  {Christy} R.~W.,  1974, \mn@doi [\prb]
  {10.1103/PhysRevB.9.5056}, \href
  {http://adsabs.harvard.edu/abs/1974PhRvB...9.5056J} {9, 5056}

\bibitem[\protect\citeauthoryear{{Jura}}{{Jura}}{2003}]{Jura_2003}
{Jura} M.,  2003, \mn@doi [\apjl] {10.1086/374036}, \href
  {http://adsabs.harvard.edu/abs/2003ApJ...584L..91J} {584, L91}

\bibitem[\protect\citeauthoryear{{Kilic}, {von Hippel}, {Leggett}  \&
  {Winget}}{{Kilic} et~al.}{2006}]{Kilic_2006}
{Kilic} M.,  {von Hippel} T.,  {Leggett} S.~K.,   {Winget} D.~E.,  2006,
  \mn@doi [\apj] {10.1086/504682}, \href
  {http://adsabs.harvard.edu/abs/2006ApJ...646..474K} {646, 474}

\bibitem[\protect\citeauthoryear{{Koester}, {G{\"a}nsicke}  \&
  {Farihi}}{{Koester} et~al.}{2014}]{Koester_2014}
{Koester} D.,  {G{\"a}nsicke} B.~T.,   {Farihi} J.,  2014, \mn@doi [\aap]
  {10.1051/0004-6361/201423691}, \href
  {http://adsabs.harvard.edu/abs/2014A%26A...566A..34K} {566, A34}

\bibitem[\protect\citeauthoryear{{Rappaport}, {Gary}, {Kaye}, {Vanderburg},
  {Croll}, {Benni}  \& {Foote}}{{Rappaport} et~al.}{2016}]{Rappaport_2016}
{Rappaport} S.,  {Gary} B.~L.,  {Kaye} T.,  {Vanderburg} A.,  {Croll} B.,
  {Benni} P.,   {Foote} J.,  2016, \mn@doi [\mnras] {10.1093/mnras/stw612},
  \href {http://adsabs.harvard.edu/abs/2016MNRAS.458.3904R} {458, 3904}

\bibitem[\protect\citeauthoryear{{Redfield}, {Farihi}, {Cauley}, {Parsons},
  {Gaensicke}  \& {Duvvuri}}{{Redfield} et~al.}{2016}]{Redfield_2016}
{Redfield} S.,  {Farihi} J.,  {Cauley} P.~W.,  {Parsons} S.~G.,  {Gaensicke}
  B.~T.,   {Duvvuri} G.,  2016, preprint, \href
  {http://adsabs.harvard.edu/abs/2016arXiv160800549R} {} (\mn@eprint {arXiv}
  {1608.00549})

\bibitem[\protect\citeauthoryear{{Shvartzvald} et~al.,}{{Shvartzvald}
  et~al.}{2016}]{Shvartzvald_2016}
{Shvartzvald} Y.,  et~al., 2016, \mn@doi [\mnras] {10.1093/mnras/stw191}, \href
  {http://adsabs.harvard.edu/abs/2016MNRAS.457.4089S} {457, 4089}

\bibitem[\protect\citeauthoryear{{Vanderburg} et~al.,}{{Vanderburg}
  et~al.}{2015}]{Vanderburg_2015}
{Vanderburg} A.,  et~al., 2015, \mn@doi [\nat] {10.1038/nature15527}, \href
  {http://adsabs.harvard.edu/abs/2015Natur.526..546V} {526, 546}

\bibitem[\protect\citeauthoryear{{Veras}}{{Veras}}{2016}]{Veras_2016}
{Veras} D.,  2016, \mn@doi [Royal Society Open Science] {10.1098/rsos.150571},
  \href {http://adsabs.harvard.edu/abs/2016RSOS....350571V} {3, 150571}

\bibitem[\protect\citeauthoryear{{Veras}, {Carter}, {Leinhardt}  \&
  {G{\"a}nsicke}}{{Veras} et~al.}{2017}]{Veras_2017}
{Veras} D.,  {Carter} P.~J.,  {Leinhardt} Z.~M.,   {G{\"a}nsicke} B.~T.,  2017,
  \mn@doi [\mnras] {10.1093/mnras/stw2748}, \href
  {http://adsabs.harvard.edu/abs/2017MNRAS.465.1008V} {465, 1008}

\bibitem[\protect\citeauthoryear{{Vogt} et~al.,}{{Vogt}
  et~al.}{1994}]{Vogt_1994}
{Vogt} S.~S.,  et~al., 1994, in {Crawford} D.~L.,  {Craine} E.~R.,  eds,
  \procspie Vol. 2198, Instrumentation in Astronomy VIII. p.~362,
  \mn@doi{10.1117/12.176725}

\bibitem[\protect\citeauthoryear{{Winn} \& {Fabrycky}}{{Winn} \&
  {Fabrycky}}{2015}]{Winn_2015}
{Winn} J.~N.,  {Fabrycky} D.~C.,  2015, \mn@doi [\araa]
  {10.1146/annurev-astro-082214-122246}, \href
  {http://adsabs.harvard.edu/abs/2015ARA%26A..53..409W} {53, 409}

\bibitem[\protect\citeauthoryear{{Xu}, {Jura}, {Dufour}  \& {Zuckerman}}{{Xu}
  et~al.}{2016}]{Xu_2016}
{Xu} S.,  {Jura} M.,  {Dufour} P.,   {Zuckerman} B.,  2016, \mn@doi [\apjl]
  {10.3847/2041-8205/816/2/L22}, \href
  {http://adsabs.harvard.edu/abs/2016ApJ...816L..22X} {816, L22}

\bibitem[\protect\citeauthoryear{{Zhou} et~al.,}{{Zhou}
  et~al.}{2016}]{Zhou_2016}
{Zhou} G.,  et~al., 2016, \mn@doi [\mnras] {10.1093/mnras/stw2286}, \href
  {http://adsabs.harvard.edu/abs/2016MNRAS.tmp.1393Z} {}

\bibitem[\protect\citeauthoryear{{Zuckerman}, {Koester}, {Reid}  \&
  {H{\"u}nsch}}{{Zuckerman} et~al.}{2003}]{Zuckerman_2003}
{Zuckerman} B.,  {Koester} D.,  {Reid} I.~N.,   {H{\"u}nsch} M.,  2003, \mn@doi
  [\apj] {10.1086/377492}, \href
  {http://adsabs.harvard.edu/abs/2003ApJ...596..477Z} {596, 477}

\bibitem[\protect\citeauthoryear{{Zuckerman}, {Melis}, {Klein}, {Koester}  \&
  {Jura}}{{Zuckerman} et~al.}{2010}]{Zuckerman_2010}
{Zuckerman} B.,  {Melis} C.,  {Klein} B.,  {Koester} D.,   {Jura} M.,  2010,
  \mn@doi [\apj] {10.1088/0004-637X/722/1/725}, \href
  {http://adsabs.harvard.edu/abs/2010ApJ...722..725Z} {722, 725}

\makeatother
\end{thebibliography}




\appendix

\section{Observing conditions during the ULTRACAM run and the reduction process}
\label{sec:ObsConditions}
\begin{figure}
\centering
\includegraphics[width=\columnwidth]{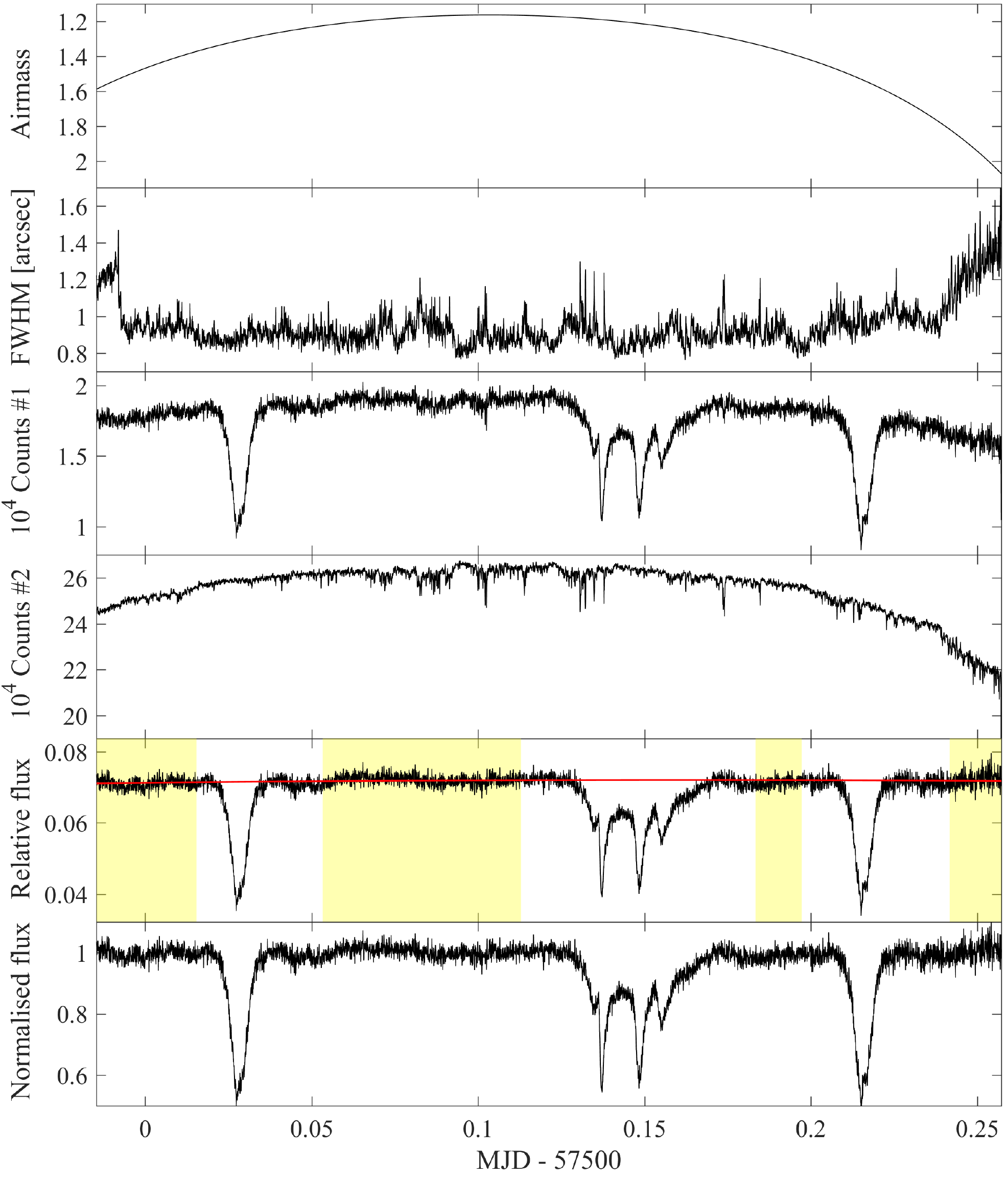}
\caption{Reduction process and observing conditions during the 2016 April 21 ULTRACAM \gb-band observations. From top to bottom: airmass; the point-source full width at half maximum; raw uncalibrated light curve of WD\,1145+017; raw light curve of the reference star, SDSS~J114825.30+013342.2, showing the atmospheric transmission; WD\,1145+017 light curve calibrated using the reference star (black), and the out-of-transit intervals (yellow) used for the parabolic fit (red); WD\,1145+017 normalised light curve.}
\label{fig:Reduction1}
\end{figure}

\begin{figure}
\centering
\includegraphics[width=\columnwidth]{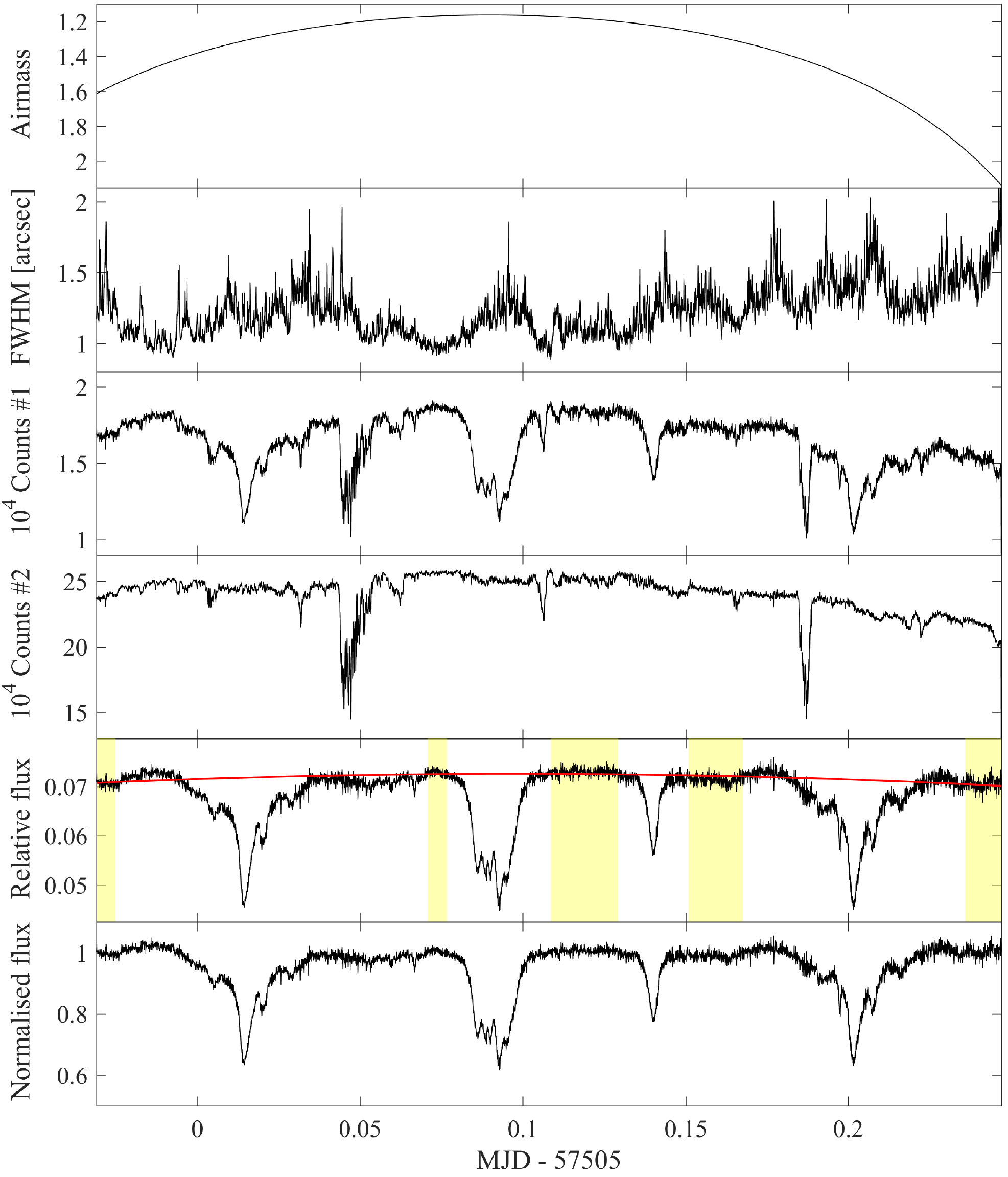}
\caption{Same as Fig.~\ref{fig:Reduction1}, for 2016 April 26. Note the sharply falling atmospheric transmission at several times (generally outside of transits, e.g. at MJDs $\sim$ 57505.05, 57505.11, and 57505.19), indicating passing clouds.}
\label{fig:Reduction2}
\end{figure}

Figs~\ref{fig:Reduction1} and \ref{fig:Reduction2} show the observing conditions during the ULTRACAM run, on 2016 April 21 and 26, as well as the reduction process described in Section~\ref{sec:Obs}. The sky transmission, reflected in the light curve of the reference star, SDSS~J114825.30+013342.2, shows some variable transmission (presumably passing clouds) on the second night, although mostly not during transits. We estimate the flux error as 1.48 times the median absolute deviation around the median, using points outside of both the transits and the passing clouds. On 2016 April 21 the errors were $\sim 4.2$, $\sim 1.9$, and $\sim 2.5$~per cent in the \ub-, \gb-, and \rb-bands, respectively, while on 2016 April 26 the errors were $\sim 2.5$~per cent (\ub-band), $\sim 1.5$~per cent (\gb-band), and $\sim 2.3$~per cent (\ib-band). The target was $\sim 33^\circ$ from the full Moon on the first night, and $\sim 93^\circ$ on the second, contributing to the better signal-to-noise ratio on the second night. The improved signal-to-noise, which reveals finer details, explains the `wiggles' seen in the out-of-transit parts of the 2016 April 26 light curve. In addition, the passing clouds contributed to the scatter in some parts of the light curve, mostly out-of-transit.

\section{Light curve evolution}
\label{sec:LC}

In parallel to our ULTRACAM observations we organised a monitoring campaign using various telescopes around the globe, in an attempt to obtain the longest continuous time coverage possible in order to follow the evolution of the light curve. A full list of the acquired light curves appears in Table~\ref{tab:Obs}. We used the observing facilities listed below. Unless otherwise mentioned, the image reduction and aperture photometry were done using the ULTRACAM pipeline \citep{Dhillon_2007}, as described in Section~\ref{sec:Obs}.

\begin{description}
\item[\textbf{Mauna Kea Observatory}]
NIR \Jb- and \Kb-band fast photometry was obtained using the WFCAM on the 3.8\,m United Kingdom Infra-Red Telescope (UKIRT) at the Mauna Kea Observatory, Hawai'i. The data were processed with pipelines from Cambridge Astronomical Survey Unit (CASU).

\item[\textbf{ARIES Devasthal Observatory}]
\rb-band photometry was obtained using the ANDOR 512 EMCCD on the 1.3\,m Devasthal Fast Optical telescope (DFOT) at the Aryabhatta Research Institute of Observational Sciences (ARIES) Devasthal Observatory in Nainital, India.

\item[\textbf{Vainu Bappu Observatory}]
\Rb-band photometry was obtained using the Princeton Pro EM CCD on the 1.3\,m J. C. Bhattacharyya Telescope (JCBT) at the Vainu Bappu Observatory (VBO) in Kavalur, India.

\item[\textbf{Wise Observatory}]
We used the FLI camera on the 0.71\,m C28 Jay Baum Rich telescope of the Wise Observatory in Israel. The photometry was obtained using the \Exopb\ filter, a high-pass filter with a cutoff around 5000\,\AA. The images were bias, dark and flat corrected using \textsc{iraf}.

\item[\textbf{Ond\v{r}ejov Observatory}]
\Rb-band photometry was obtained using the 0.65\,m D65 telescope at the Ond\v{r}ejov Observatory in the Czech Republic. The image reduction (dark subtraction and sky flat correction) and aperture photometry were done using the Ond\v{r}ejov custom-made software \textsc{Aphot}. A fixed aperture-size of 7~pixels (7.3\,arcsec) was used. The light curves were calibrated using the three brightest non-variable stars in the field.

\item[\textbf{LCO global network}]
We used the 1\,m and 0.4\,m Las Cumbres Observatory (LCO) global telescope network to obtain unfiltered photometry. The 1\,m observations were performed using the kb70 and kb78 SBIG CCDs on the cpt1m010 and lsc1m005 telescopes in Sutherland (CPT), South Africa, and in Cerro Tololo (LSC), Chile. The images were reduced by the BANZAI pipeline.
The 0.4\,m observations were performed using the kb27 and kb84 SBIG CCDs on the ogg0m406 and coj0m405 telescopes in Haleakala (OGG), Hawai'i, and in Siding Spring (COJ), Australia. The images were reduced by the ORAC-DR pipeline.

\item[\textbf{UC Observatory}]
We obtained \ib-band photometry using the SBIG STL-1001 3 CCD Camera on the 0.4\,m Tololo 40 telescope at the Pontificia Universidad Cat\'{o}lica de Chile Observatory (UCO) in Lo Barnechea, Chile.
\end{description}

\begin{table*}
\caption{List of observations}
\label{tab:Obs}
\begin{center}
\begin{tabular}[t]{l l l p{4.5cm} l l l}
\hline
Observatory &
Telescope size [m] &
Date [UT] &
Time [UT] &
Filter &
Exposure time [s] &
Cadence [s] \\
\hline
UKIRT & 3.80 & 2016-04-16 & 06:14-08:36, 08:46-10:47, 11:00-11:12 & \Jb\ & 10 & 12 \\ 
ARIES & 1.30 & 2016-04-21 & 14:04-14:37, 15:38-18:32, 18:55-20:04 & \rb\ & 180 & 180 \\ 
Ond\v{r}ejov & 0.65 & 2016-04-21/22 & 19:07-00:16 & \Rb\ & 180 & 186 \\ 
{\bf La Silla} & {\bf 3.58} & {\bf 2016-04-21/22} & {\bf 23:39-06:10} & {\bf \ub, \gb, \rb} & {\bf 10, 5, 5} & {\bf 10, 5, 5} \\ 
UKIRT & 3.80 & 2016-04-22 & 06:19-10:47 & \Jb\ & 10 & 12 \\ 
UKIRT & 3.80 & 2016-04-23 & 06:11-07:59, 08:06-09:54 & \Jb\ & 10 & 12 \\ 
ARIES & 1.30 & 2016-04-23 & 13:56-14:00, 14:59-20:04 & \rb\ & 180 & 180 \\ 
ARIES & 1.30 & 2016-04-23 & 14:08-14:10, 14:34-14:54 & \rb\ & 120 & 120 \\ 
ARIES & 1.30 & 2016-04-23 & 14:16-14:32 & \rb\ & 90 & 90 \\ 
LCO CPT & 1.00 & 2016-04-23 & 17:31-23:43 & clear & 60 & 80 \\ 
Wise & 0.71 & 2016-04-23 & 19:39-22:47 & \Exopb\ & 60 & 83 \\ 
ARIES & 1.30 & 2016-04-24 & 13:51-18:41 & \rb\ & 90 & 90 \\ 
Wise & 0.71 & 2016-04-24 & 17:23-21:50 & \Exopb\ & 60 & 83 \\ 
ARIES & 1.30 & 2016-04-24 & 18:44-20:12 & \rb\ & 180 & 180 \\ 
LCO OGG & 0.40 & 2016-04-25 & 06:02-08:53 & clear & 270 & 280 \\ 
UKIRT & 3.80 & 2016-04-25 & 06:04-08:17, 08:27-10:41 & \Jb\ & 10 & 12 \\ 
LCO COJ & 0.40 & 2016-04-25 & 09:02-09:35 & clear & 270 & 279 \\ 
ARIES & 1.30 & 2016-04-25 & 13:52-18:23 & \rb\ & 60 & 60 \\ 
VBO & 1.30 & 2016-04-25 & 15:18-17:50 & \Rb\ & 600 & 600 \\ 
Wise & 0.71 & 2016-04-25 & 17:30-21:40, 21:55-23:35 & \Exopb\ & 60 & 83 \\ 
ARIES & 1.30 & 2016-04-25 & 18:26-20:00 & \rb\ & 120 & 120 \\ 
Ond\v{r}ejov & 0.65 & 2016-04-25 & 19:18-23:19 & \Rb\ & 180 & 186 \\ 
VBO & 1.30 & 2016-04-26 & 14:01-19:37 & \Rb\ & 900 & 900 \\ 
ARIES & 1.30 & 2016-04-26 & 14:05-19:27, 19:33-20:17 & \rb\ & 60 & 60 \\ 
Ond\v{r}ejov & 0.65 & 2016-04-26 & 19:37-23:11 & \Rb\ & 180 & 186 \\ 
{\bf La Silla} & {\bf 3.58} & {\bf 2016-04-26/27} & {\bf 23:15-05:55} & {\bf \ub, \gb, \ib} & {\bf 10, 5, 5} & {\bf 10, 5, 5} \\ 
LCO LSC & 1.00 & 2016-04-26/27 & 23:45-23:59, 00:08-00:38, 00:54-00:56, 01:12-01:22, 01:33-01:53, 02:13-03:26, 03:35-04:44, 04:56-05:16 & clear & 60 & 84 \\ 
UCO & 0.40 & 2016-04-27 & 00:22-01:16 & \ib\ & 180 & 196 \\ 
UCO & 0.40 & 2016-04-27 & 01:18-02:57 & \ib\ & 120 & 134 \\ 
LCO OGG & 0.40 & 2016-04-27 & 06:02-08:53 & clear & 270 & 280 \\ 
LCO COJ & 0.40 & 2016-04-27 & 09:02-10:53, 12:07-12:53 & clear & 270 & 279 \\ 
VBO & 1.30 & 2016-04-27 & 14:23-16:56, 19:13-19:43 & \Rb\ & 900 & 900 \\ 
Ond\v{r}ejov & 0.65 & 2016-04-27 & 19:46-20:54, 21:20-23:11, 23:45-23:55 & \Rb\ & 180 & 186 \\ 
VBO & 1.30 & 2016-04-28 & 13:54-19:17 & \Rb\ & 600 & 600 \\ 
LCO OGG & 0.40 & 2016-04-29 & 07:02-07:54, 10:02-10:54 & clear & 270 & 281 \\ 
UKIRT & 3.80 & 2016-05-02 & 05:42-08:10, 08:18-10:45 & \Kb\ & 10 & 34 \\ 

\hline
\end{tabular}
\end{center}
\end{table*}

Combining all our light curves with those kindly provided by \citet{Gary_2017} results in a long-baseline light curve of about half a year, with denser temporal coverage during our week-long monitoring campaign. We converted the UT timestamps to Barycentric Julian Dates (BJD) in Barycentric Dynamical Time (TDB) using the tool provided by \citet{Eastman_2010}\footnote{\url{http://astroutils.astronomy.ohio-state.edu/time/utc2bjd.html}}.

\begin{figure}
\includegraphics[width=\columnwidth]{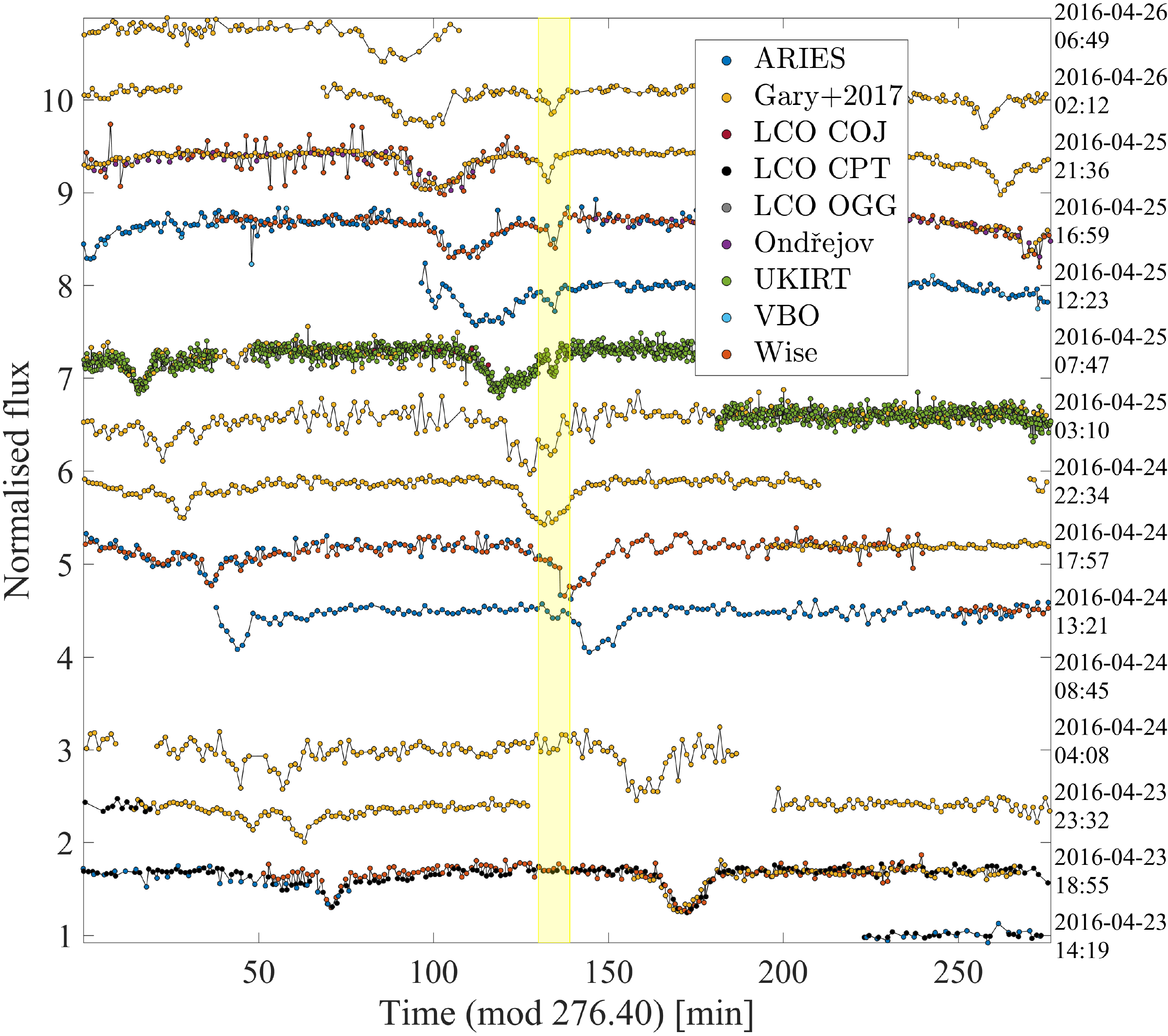}
\caption{Part of the combined light curve, folded over the `B'-period (276.40\,min). Each cycle is vertically shifted by 0.6 for clarity. The highlighted region marks the `B'-dip appearance, around 2016 April 24 15:36 (BJD TDB).}
\label{fig:BDip}
\end{figure}

Although the original \textit{K2} observations showed six stable periods between $\sim4.5-5$\,h, denoted `A'-`F' \citep{Vanderburg_2015}, only the shortest, `A'-period ($\sim 4.5$\,h), was detected in ground-based follow-up observations. \citet{Gary_2017} reported the first ground-based detection of a longer-period dip feature, corresponding to the `B'-period ($\sim 4.6$\,h) of \citet{Vanderburg_2015}, starting on 2016 April 26. During our own monitoring campaign, three distinctive dip groups were identified, each sharing a common period and a close location in phase. Two of these groups, denoted `A1' and `A2' in Figs~\ref{fig:Color20160421} and \ref{fig:Color20160426}, share a period close to the original \textit{K2} `A'-period: $269.470\pm0.005$\,min and $269.9\pm0.3$\,min, respectively. These are the groups denoted as `G6121' and `G6420' in \citet{Gary_2017}, indicating their first detection date (2016 January 21 and 2016 April 20, respectively). The third group, denoted as `B' in Fig.~\ref{fig:Color20160426}, corresponds to the `B-dip' group of \citet{Gary_2017}, with a period of $276.40\pm0.03$\,min, close to the original \textit{K2} `B'-period. Combining the light curves from our monitoring campaign with those provided by \citet{Gary_2017}, we were able to further constrain the appearance of the `B' dip to 2016 April 24 (see Fig.~\ref{fig:BDip}). Since the light curve coverage was not complete on this date, we cannot identify the exact appearance time of the `B' dip. It was first detected around 2016 April 24 15:36 (BJD TDB), and might have been present already on 2016 April 24 06:22 (BJD TDB).

\begin{figure}
\includegraphics[width=\columnwidth]{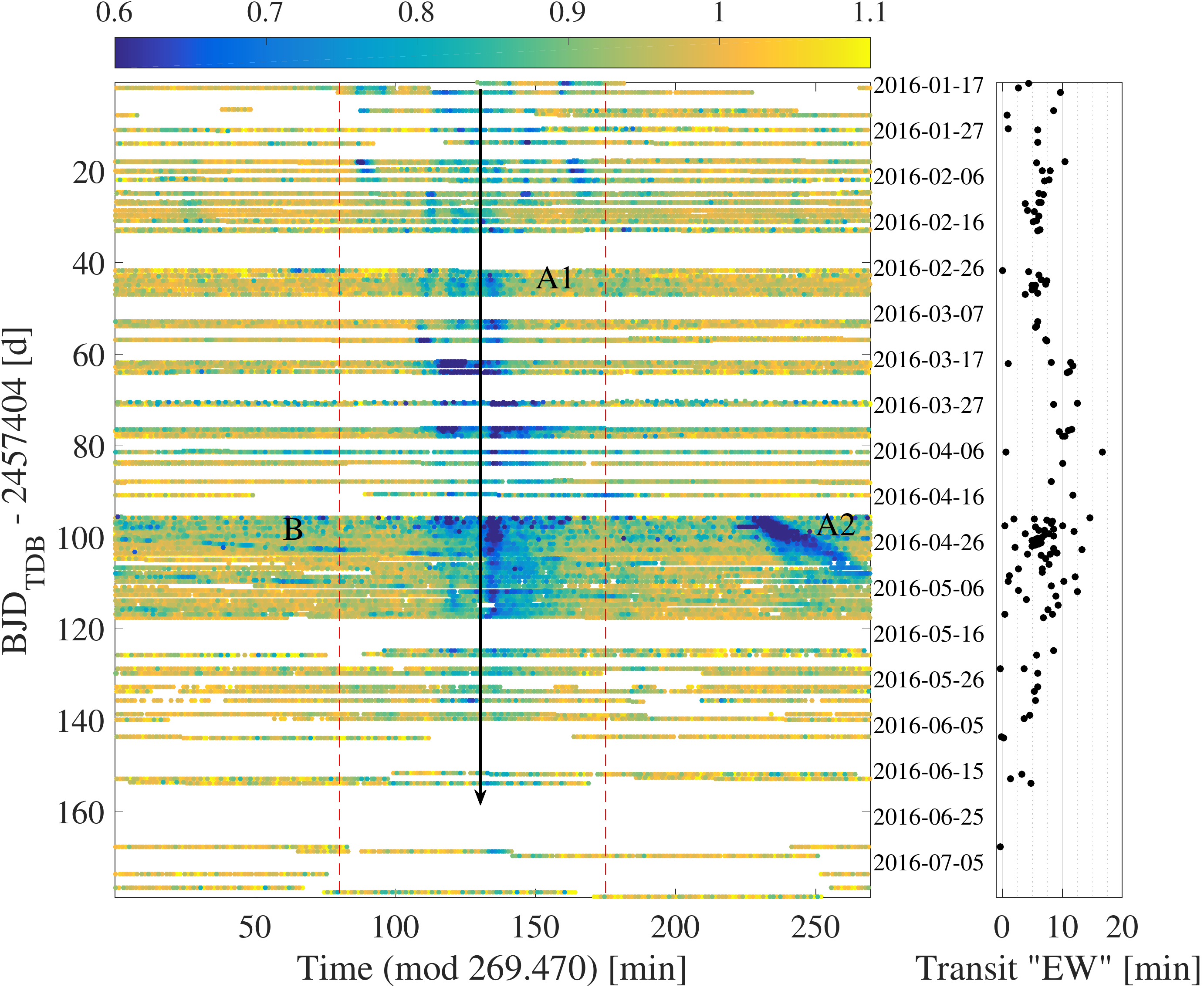}
\caption{Left: `Raw waterfall' plot of the combined light curve, folded over the `A1' period (269.470\,min) and coloured by amplitude (from blue to yellow, with normalised flux ranging from 0.6 to 1.1). There are several dips sharing roughly the same period, and lasting a few months. The black vertical arrow represents a constant period. Transits with different orbital periods (`A2' and `B') appear as diagonal patterns. Right: transit `EW'. The dashed red lines on the left panel mark the EW integration limits.}
\label{fig:LC_A1}
\end{figure}

\begin{figure}
\includegraphics[width=\columnwidth]{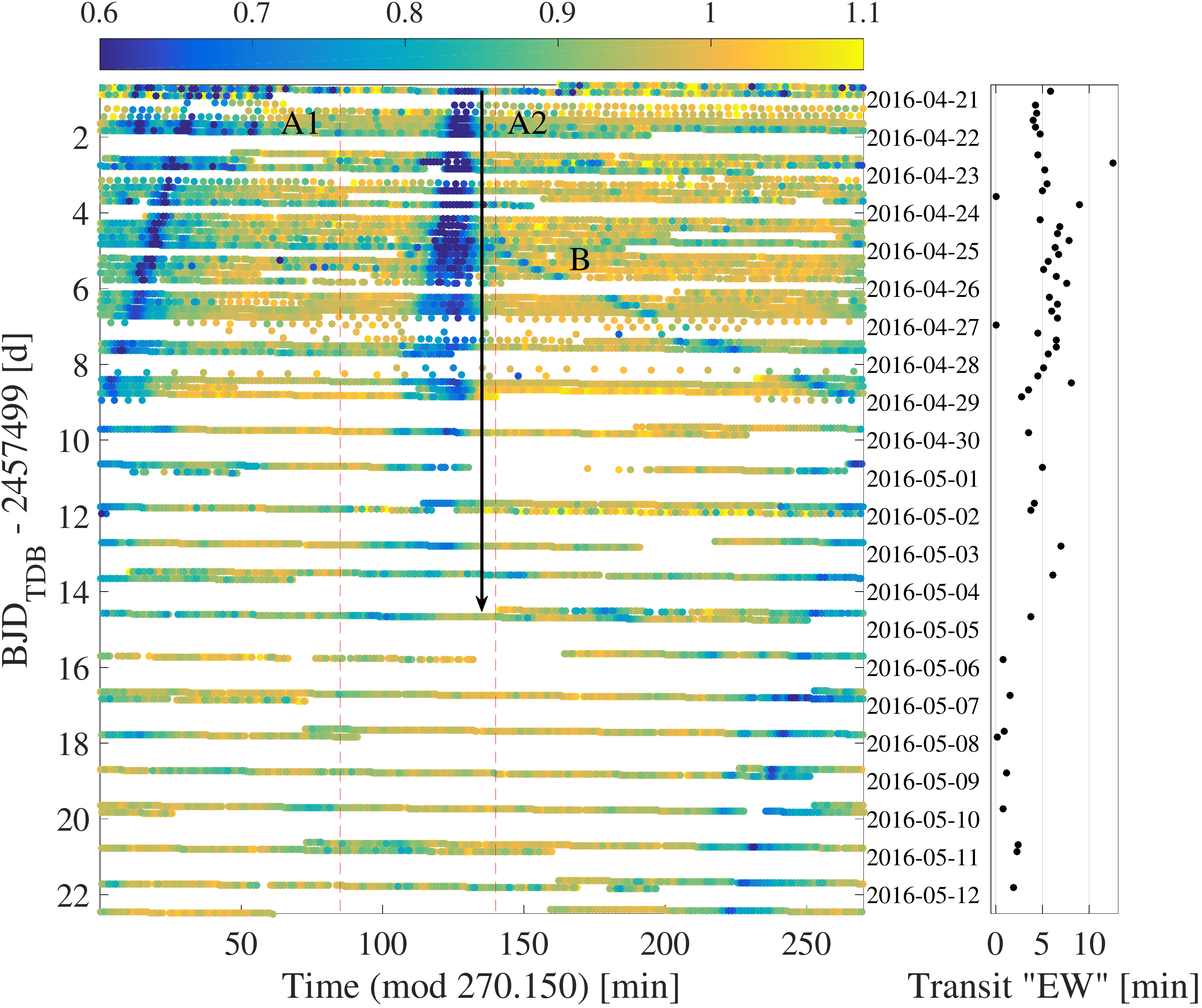}
\caption{Same as Fig.~\ref{fig:LC_A1}, folded over the `A2' period (270.15\,min). This group of dips was first detected by \citet{Gary_2017} on 2016 April 21. Compared to the black vertical arrow representing a constant period, the data indicate a decrease in the dip group's period.}
\label{fig:LC_A2}
\end{figure}

\begin{figure}
\includegraphics[width=\columnwidth]{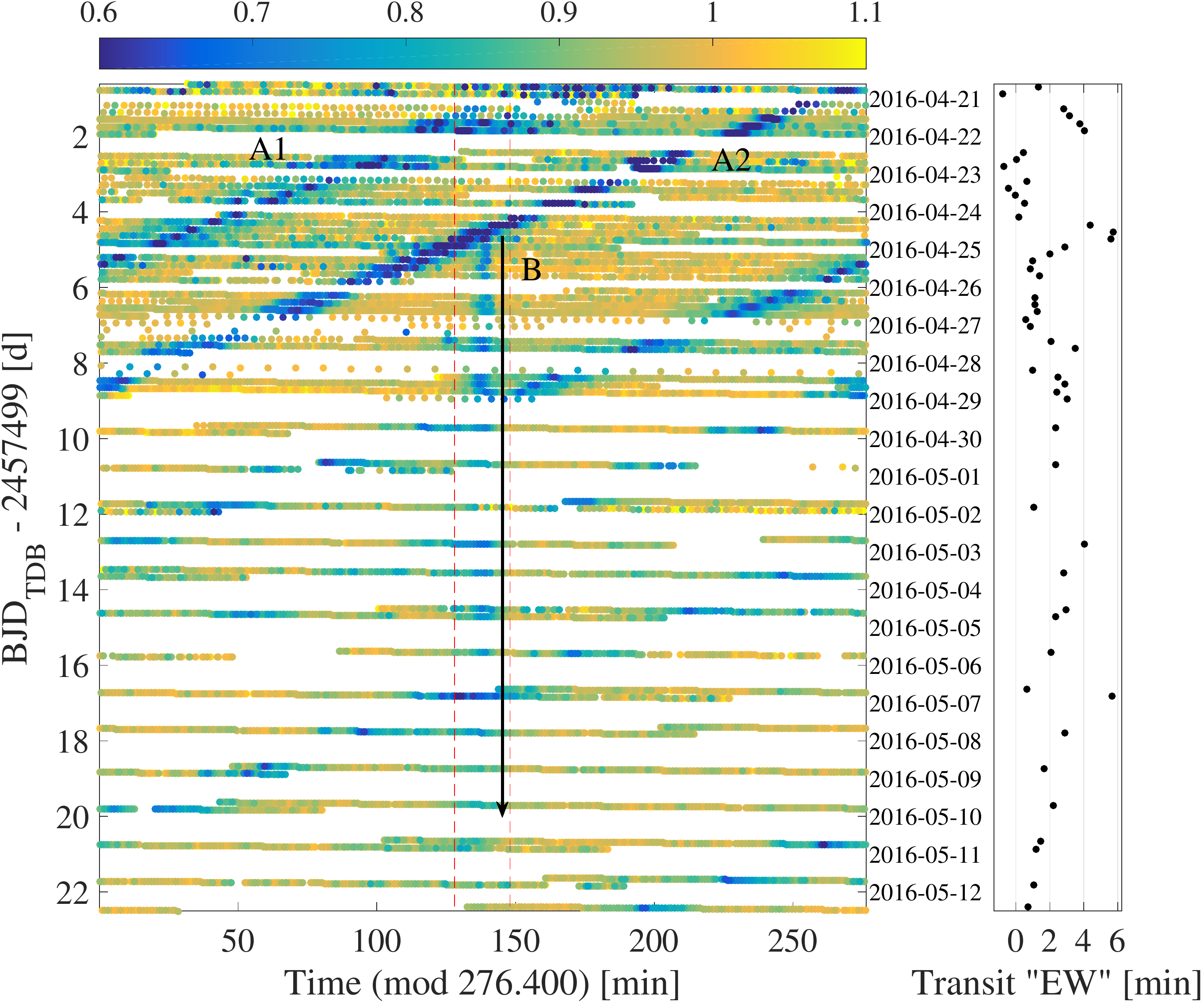}
\caption{Same as Fig.~\ref{fig:LC_A1}, folded over the `B' period (276.40\,min).}
\label{fig:LC_B}
\end{figure}

Figs~\ref{fig:LC_A1}, \ref{fig:LC_A2}, and \ref{fig:LC_B} present a `raw waterfall' plot of the combined light curve, for the `A1', `A2', and `B' period, respectively. In a `raw waterfall' the light curve is folded over the period, plotted against the actual timing of its sample points, and coloured by the normalised flux (the bluer the lower). Thus, a raw waterfall plot of a dip feature with a constant period would appear as a straight vertical blue line. A non-vertical straight line pointing to the right (left) would imply that the plot is folded over a period shorter (longer) than the actual period of the dip. A curved line pointing to the right (left) would imply a gradually increasing (decreasing) period. The periods mentioned above were measured manually `by eye' using the raw waterfall plots. The period uncertainty represents the range of periods either yielding a relatively straight vertical line, or corresponding to different parts of the slope in case of a curved line.

Fig.~\ref{fig:LC_A1}, which features the dip group of period `A1', shows a stable period over about half a year. Fig.~\ref{fig:LC_A2} demonstrates what appears to be the `A2' cloud breaking up, as it gets shallower and wider in time. This time we see a decrease in the group elements' period over a few days, as it deviates from the reference vertical straight arrow in the plot, indicating a decreasing orbital separation. The `A2' group seems to vanish after a couple of weeks. Finally, Fig.~\ref{fig:LC_B} shows the appearance (and apparently disappearance) of the `B'-dip. The `B'-period does not seem to deviate much. All these figures (\ref{fig:LC_A1}-\ref{fig:LC_B}) share the same colour scale (from blue to yellow, with normalised flux ranging from 0.6 to 1.1). Following \citet{Gary_2017}, we have calculated the `transit EW' for the `A2'- and `B'-dips, defined as the area between the out-of-transit unity-normalised level and the light curve, integrated over a specific time interval.\footnote{This definition differs from the one used by \citet{Gary_2017}, who used a dimensionless EW calculated using fitted parameters.} This quantity reflects the total obscured area. The transit EW appears along the raw waterfall plot in Figs~\ref{fig:LC_A1}-\ref{fig:LC_B}. While for the `A1'-dip, which has a lifetime of more than half a year, and is consisted of a few dip-features, it was harder to define the integration limits, for the shorter-lived `A2'- and `B'-dips it was more straight-forward. The transit EW seems to be relatively constant during the lifetimes of both the `A2'- and `B'-dips, independent of the dip's shape, with outliers resulting from temporal overlap with other dips, or due to an insufficient number of samples within the relevant time interval. The total `A2'-dip EW appears to maintain a $\sim 5$\,min value for a couple of weeks, while its cloud breaks-up and spreads. The `B'-dip also maintains a relatively constant EW of $\sim 2$\,min for a little more than two weeks. In this case, the spread is less evident.

\begin{figure}
\centering
\includegraphics[width=0.8\columnwidth]{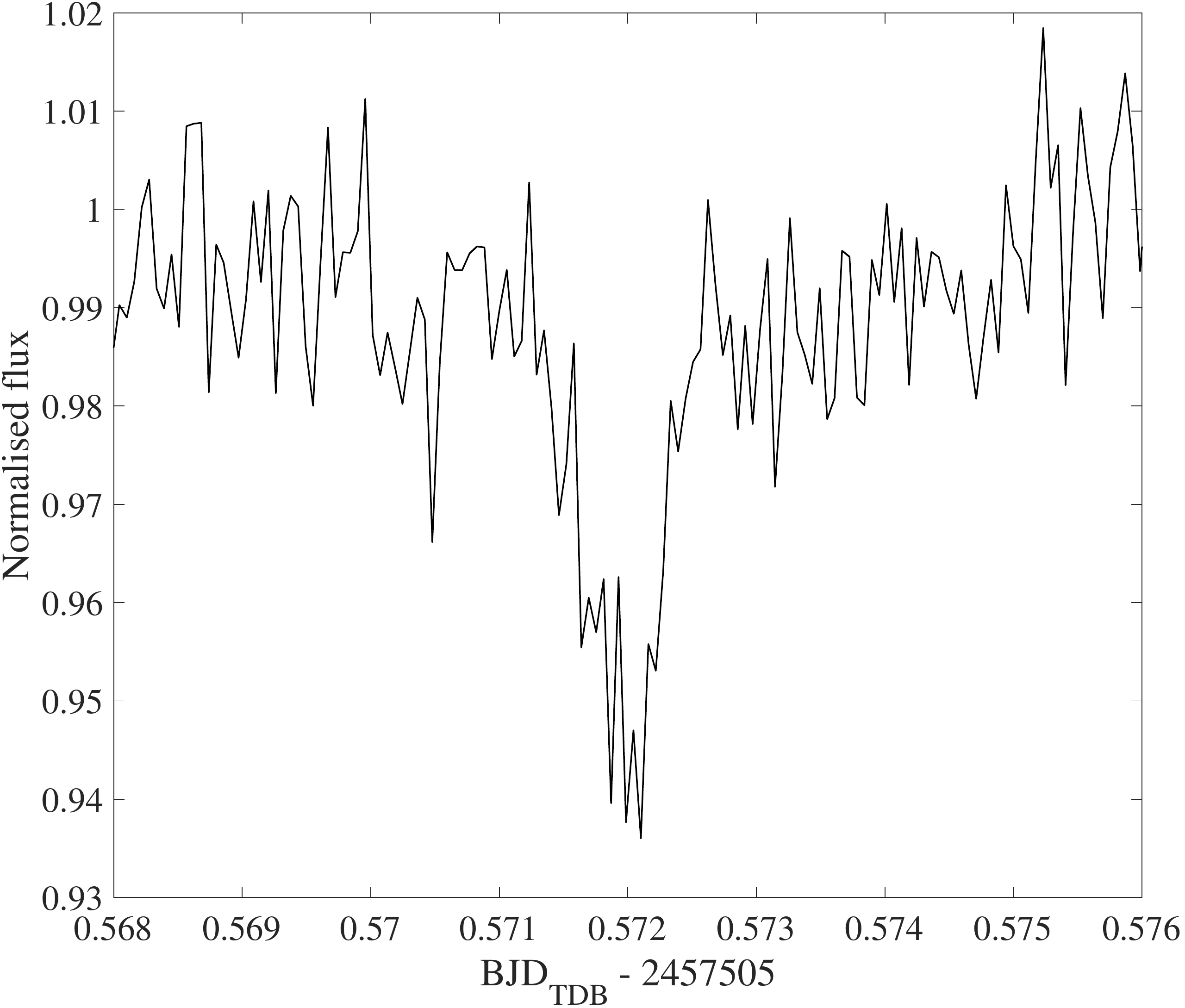}
\caption{Part of our 2016 April 26 unbinned \gb-band ULTRACAM light curve, showing an example of a smaller dip feature.}
\label{fig:Fine}
\end{figure}

The high signal-to-noise ratio ULTRACAM light curve from 2016 April 26 (when the object was far enough from the moon), shows some finer details, undetectable by smaller telescopes. Fig.~\ref{fig:Fine} shows an example of such a transit, shorter ($\sim 1.5$\,min) and shallower ($\sim 5$~per cent) than usually detected. Unfortunately, we were not able to observe the target for more than one cycle under these conditions, which prevents us from estimating the exact period of this feature. In addition, there is some suggestion of an increase of flux preceding the two `A1' transits on 2016 April 26 (see Fig.~\ref{fig:Color20160426}), possibly the result of scattering of the WD flux from the dust cloud prior to the transit. However, there is a good chance that these are calibration artefacts resulting from the difficulty in defining the out-of-transit parts of the light curve, and they require further investigation. If this flux excess is real, it might help to constrain the dust grain-size distribution \citep[e.g][]{Brogi_2012}. As can be seen in Fig.~\ref{fig:Color20160426}, the out-of-transit area around the `A1' group is quite ambiguous, indicating the presence of obscuring material, probably originating from the same group. This is also evident in Fig.~\ref{fig:LC_A1} by the light blue regions surrounding the darker, deeper, transits of the `A1' group.

%
%


\end{CJK}

\bsp	
\label{lastpage}
\end{document}